\documentclass[journal]{IEEEtran}
\usepackage{cite}

\ifCLASSINFOpdf
   \usepackage[pdftex]{graphicx}
   \graphicspath{{../pdf/}{../jpeg/}}
   \DeclareGraphicsExtensions{.pdf,.jpeg,.png}
\else
\fi

\usepackage{amsmath}
\usepackage{algorithmic}
\usepackage{color}
\usepackage[modulo,switch]{lineno}  
\usepackage{stfloats} 
\usepackage{array}
\usepackage{multirow}

\begin{document}
%

\title{Low-light Image Enhancement Using the Cell Vibration Model}

\author{Xiaozhou~Lei,~Zixiang~Fei*,~Wenju~Zhou,~Huiyu~Zhou~and~Minrui~Fei
\thanks{The work was supported by the Natural Science Foundation of China (No. 61877065), Key Project of Science and Technology Commission of Shanghai Municipality (No. 19510750300), and 111 Project (No. D18003).}
\thanks{X. Lei, M. Fei and W. Zhou are with the Shanghai Key Laboratory of Power Station Automation Technology, School of Mechatronical Engineering and Automation, Shanghai University, Shanghai 200444, China (e-mail: sigmoid@qq.com; mrfei@staff.shu.edu.cn; zhouwenju@shu.edu.cn).}
\thanks{Z. Fei is with the School of Computer Engineering and Science, Shanghai University, Shanghai 200444, China (e-mail: zxfei@shu.edu.cn).}
\thanks{H. Zhou is with the School of Computing and Mathematical Sciences, University of Leicester, Leicester, LE1 7RH, United Kingdom (e-mail: hz143@leicester.ac.uk).}}

\markboth{IEEE TRANSACTIONS ON MULTIMEDIA,~Vol~1, No.~1, May~2022, ACCEPTED VERSION, NOT Published VERSION}%
{Shell \MakeLowercase{\textit{et al.}}: Bare Demo of IEEEtran.cls for IEEE Journals}

\maketitle

\begin{abstract}
Low light very likely leads to the degradation of an image's quality and even causes visual task failures. Existing image enhancement technologies are prone to overenhancement, color distortion or time consumption, and their adaptability is fairly limited. Therefore, we propose a new single low-light image lightness enhancement method. First, an energy model is presented based on the analysis of membrane vibrations induced by photon stimulations. Then, based on the unique mathematical properties of the energy model and combined with the gamma correction model, a new global lightness enhancement model is proposed. Furthermore, a special relationship between image lightness and gamma intensity is found. Finally, a local fusion strategy, including segmentation, filtering and fusion, is proposed to optimize the local details of the global lightness enhancement images. Experimental results show that the proposed algorithm is superior to nine state-of-the-art methods in avoiding color distortion, restoring the textures of dark areas, reproducing natural colors and reducing time cost. The image source and code will be released at https://github.com/leixiaozhou/CDEFmethod.
\end{abstract}

\begin{IEEEkeywords}
Low light, Image enhancement, Cell vibration model, Guided filtering, Image fusion.
\end{IEEEkeywords}

\IEEEpeerreviewmaketitle

\section{Introduction}
\IEEEPARstart{W}{ith} increasing demands for video recordings {\color{blue}created from} night scenes, where images are acquired without sufficient exposure, people must deal with problems such as low lightness, low contrast and increasing image noise. Image content may be missing or obscured due to darkness. Therefore, suitable image enhancement technologies are being developed to reveal the image information hidden in the dark \cite{b1,b2}.

In the field of low light image enhancement, one of the classic and extensively {\color{blue}utilized} methods is histogram-based techniques. {\color{blue}Histogram equalization (HE) \cite{b7} is a fast yet simple method that adjusts the} corresponding distributions of individual RGB channels in a color image. Problems such as {\color{blue}overenhancement}, color distortion, and evident noise in dark areas cannot be avoided due to {\color{blue}a lack of color correlation constraints} across channels. To cope with these shortcomings, several HE-based methods have been proposed, such as CLAHE \cite{b8}, CVC \cite{b9}, and MLHE \cite{b10,b11}. Celik and Tjahjadi \cite{b9} reported that the input image contrast can be enhanced by constructing a two-dimensional histogram based on the neighborhood {\color{blue}relationships} of image pixels. The disadvantage of the method is that color restoration still needs to be improved \cite{b12,b13,b14}. Caselles \emph{et al.}\cite{b10} proposed a local contrast enhancement algorithm for shape preservation. The disadvantage is that the image noise is also amplified \cite{b11}.

Other widely influential methods include model-based approaches, such as the Retinex model \cite{b2,b15,b16,b17,b18,b19,b20,b21,b46,b47,b64}, tone mapping method \cite{b22,b23,b24,b25,b26,b48}, physical lighting model \cite{b14,b27,b28,b29,b30}, and frequency-domain methods\cite{b74,b75,b76}. For the {\color{blue}retinex} model, the LIME algorithm proposed by Guo \cite{b2} has attracted much attention. By building a smooth structure perception model, the illumination consistency can be improved, and finally, a well-structured illumination map is obtained. Gu \emph{et al.} \cite{b21} proposed a fractional-order variational model based on {\color{blue}retinex} to obtain appropriate illumination estimation results. Hao \emph{et al.} \cite{b64} proposed a {\color{blue}semidecoupled} Gaussian total variation model to estimate the illumination and reflectance layers. Singh \emph{et al.} \cite{b70} proposed a {\color{blue}texturewise} adaptive gamma correction method and a quintile-based reflectance calculation method to enhance textures and improve unbalanced/nonuniform illumination. The advantage of this method is that texture enhancements {\color{blue}are implemented} in a non-iterative way. However, the calculation is time-consuming. Notably, retinex-based methods often use gamma transformations to readjust illumination estimation results.

For tone mapping methods, Ahn and Keum \cite{b24} proposed {\color{blue}enhancing} dark images using a two-step method {\color{blue}that included} global mapping and local color adjustment. This method can retrieve image details and has less computational cost. Srinivas \emph{et al.} \cite{b48} proposed an adaptive sigmoid transformation function to adjust lightness, {\color{blue}and used a Laplace filter to enhance the edge strength.} Li \emph{et al.} \cite{b67} presented an algorithm that adaptively adjusts the global contrast of gray images using the bilateral gamma adjustment function and particle swarm optimization. The advantage is {\color{blue}its reduction of} the influence of uneven illumination on the image quality but it is limited to considering only gray images rather than color images. In addition, because the final enhanced image is obtained by iterative calculations, it is very time-consuming.
For physical lighting models, Jiang and Yao \cite{b27} constructed a foggy image by reversing the pixels of the dark image and then used the dark channel algorithm to remove the noise to improve the contrast of the dark image. Based on the illumination-reflection model and multiscale theory, Wanga and Chen \cite{b14} presented a color correction method using a {\color{blue}nonlinear} transformation function. 

For frequency-domain methods, Zhang et al. \cite{b74} enhanced the local contrast information of uneven-illuminated images by combining the homomorphic filtering method with discrete cosine transformation (DCT). {\color{blue}The blocking effect} of the homomorphic filtering method is reduced in the spatial domain by using gradient information and DCT features. Kawasaki et al. \cite{b75} used the LL component of wavelet expansion to estimate the illumination component and used 2-level wavelet decomposition and reconstruction methods to improve the accuracy of the illumination components. This method improves the performance of the traditional MSR methods, but it is time consuming. AAMIR et al. \cite{b76} proposed a low light level enhancement method using wavelet transform. {\color{blue}A dual-tree} complex wavelet is used to decompose V-channel images to obtain high- and low-frequency components. The low-frequency component is enhanced to improve the brightness of the image, and the noise of the high-frequency component is suppressed to strengthen the texture features.

Deep learning technology offers potential solutions to the field of low-light image enhancement. Lore \emph{et al.} \cite{b31} proposed a superimposed sparse denoising {\color{blue}autoencoder} (SSDA). This method can adaptively enhance the lightness of natural low-light images, and {\color{blue}recognizes} the characteristics of the images. Chen \emph{et al.} \cite{b32} established a short exposure and low light image dataset, including the corresponding long exposure reference image. Using this dataset, they developed a dark image enhancement method based on the end-to-end full convolution network.
Zhang \emph{et al.} \cite{b33} proposed a deep network for kindling the darkness (KinD), which trained multiple exposure image pairs to estimate the appropriate illumination and reflectance.
Lim and Kim \cite{b66} proposed a deep stacked Laplacian restorer (DSLR) that uses different levels of Laplacian pyramids to guide the encoder in restoring local details and global illumination. Deep learning has driven advances in image enhancement but the disadvantage is that it requires very high computing power on the device, and these algorithms are usually time consuming. For small devices such as mobile phones, faster computing methods need to be explored.

{\color{blue}The exposure} fusion method fuses two or more exposure images to obtain a {\color{blue}visually harmonious image}. Li \emph{et al.} \cite{b53} proposed a multimode image fusion method based on guided filtering, which accelerates the fusion process. {\color{blue}The disadvantage is that the two images with little indifference exposure will have severe color distortion after fusion.} Fu \emph{et al.} \cite{b51} used a guided filter to estimate the illumination and then designed weights to fuse the lightness-improved and contrast-enhanced images. Ying \emph{et al.} \cite{b52} proposed an exposure fusion framework based on a camera response model and light estimation. A well-exposed fused image is obtained with appropriate exposure and a fusion weight {\color{blue}was} designed using light estimation techniques. Singh \emph{et al.} \cite{b69} proposed a fusion framework to fuse the channel results of piecewise gamma correction and histogram equalization and used particle swarm optimization (PSO) to find the appropriate multichannel fusion coefficients. The simplicity of the fusion framework is sought, but the inefficiency of the PSO algorithm limits its ability to perform real-time computations.
Chang \emph{et al.} \cite{b65} proposed a long-short-exposure fusion network (LSFNet) to solve the problems caused by fusion, such as high noise, motion blur and color distortion.

In this paper, we propose a fast and effective method for low-light image enhancement. The proposed solution has three main contributions.
\begin{enumerate}   
	\item We analyze the widespread phenomenon of cell vibration and establish an energy model of the stimulus intensity by using the mechanism of vibration, which quantitatively describes the relationship between stimulus intensity and energy in the process of cell photothermal transformation.

	\item We propose a new global lightness enhancement model that integrates the energy model with the gamma correction model, which improves the deficiency of the gamma correction model. Then, we find the statistical property of lightness and gamma intensity that both the proposed model and the traditional gamma model have, which is helpful to automatically adjust the gamma intensity parameters.
	
	\item We propose a local fusion strategy to improve the results of global lightness enhancements and to restore the details of the defect areas. \end{enumerate}

The remainder of the paper is {\color{blue}organized in} as follows: In Section \ref{sec:method}, we fully discuss the proposed energy model, global lightness enhancement methods, and local fusion strategy. Section \ref{sec:exp} shows various experimental analyses and results. Our conclusions are given in Section \ref{sec:conc}.

\section{Methodology}
\label{sec:method}
\subsection{Energy Model}
Light transfers energy by independent photons \cite{b34}, and photoelectric and photothermal effects occur when the photons are absorbed by retinal cells. Energy is converted into electricity and heat. This triggers the movement of ions on the retinal cells, which rely on electrical signals generated by ion movements to produce images \cite{b56}.
This paper focuses on the cell changes caused by the photothermal conversion process and attempts to quantify the portion of the energy that is converted to heat. 

Retinal cells consist of elastic cell membranes and liquid cytoplasm. When photons are absorbed by the retinal cell, energy is transferred from photons to the retinal cell, causing the retinal cell temperature to rise. Therefore, the liquid cytoplasm expands when heated, and the elastic cell membrane expands synchronously. Without subsequent energy supplies, the volume of liquid cytoplasm shrinks, and the cell membrane returns to its initial undisturbed state after the cell temperature drops. Based on this dynamic behavior of expansion and contraction, a free vibration model of a damped system is constructed using mechanical vibration theory. Correspondingly, cell membranes with elasticity are {\color{blue}considered} elastic elements, and liquid {\color{blue}cytoplasm} with energy absorption and shock reduction are {\color{blue}considered} damping elements. Because photons cause motion, photons are {\color{blue}considered} objects, and the stimulus intensity of photons on retinal cells is {\color{blue}considered mass. The ability of cells to generate nano-vibrations as a result of heat is ubiquitous and considered a basis for the existence of life signals \cite{b54,b55}.}

Assuming the amplitude of cell vibration is $s(t)$, based on the theory of mechanical vibration, we assume that cell membranes that can expand and shrink dynamically are elastic elements with a {\color{blue}stiffness of $k$}, and the cytoplasm that can absorb energy and reduce vibration is a damper with a viscous damping coefficient of $c$. The stimulation intensity caused by a single photon is set as the equivalent mass of weight $M$. Therefore, the displacement $s(t)$ produced by a single vibration of the cell membrane is:
\begin{equation}\label{eq:eq2}
	\ddot{s}(t)+2 \beta \dot{s}(t)+\omega_{0}^{2} s(t)=0
\end{equation}
where $\omega_{0}^{2}=\frac{k}{M}$, $\beta=\frac{c}{2 M}$, and $M>0$. $\beta$ is a viscous damping factor or damping rate, which is a dimensionless parameter. $\omega_{0}$ is the natural frequency of the system. When the stimulus intensity is determined, the natural frequency is only determined by the parameters of the system itself, which is independent of the external stimulus, initial conditions, \emph{etc.} {\color{blue}in addition to} the current stimulus intensity. For the selection of a general solution about $s(t)$, we consider the case of critical damping ($\beta=\omega_{0}$). 

{\color{blue}If there is no subsequent energy supplement, the retinal cells will return to the initial stability due to energy dissipation.} In this case, the system vibrates only once and returns to the static balance position in a short time. {\color{blue}The solution that satisfies the process of cell vibration is:}
\begin{equation}\label{eq:eq3}
	s(t)=e^{-\beta t}\left(c_{1}+c_{2} t\right)
\end{equation}
where $c_{1}$ and $c_{2}$ are two integral constants determined by the starting condition of a motion. Because the energy carried by a single photon currently absorbed is the {\color{blue} constant, which means that the stimulus intensity is constant over time, that is} $\frac{d M}{d t}=0$, the energy required for the vibration process is:
\begin{equation}\label{eq:eq4}
	\varepsilon=\int_{0}^{+\infty}\left(v^{2} \frac{d M}{d t}+M v \frac{d v}{d t}\right) d t=\left.\frac{1}{2} M v^{2}\right|_{0} ^{+\infty}
\end{equation}
where, because velocity $v$ and displacement $s$ have a differential relationship, i.e.

\begin{equation}\label{eq:eq22}
	v=\frac{d s}{d t}=c_{2}e^{-\beta t}-\beta e^{-\beta t}\left(c_{1}+c_{2} t\right)
\end{equation}

Thus, Eqs.\eqref{eq:eq4} and \eqref{eq:eq22} can help us derive the expression of $\varepsilon$.

\begin{equation}\label{eq:eq23}
	\varepsilon=c_{1} c_{2} \sqrt{k M}-\frac{1}{2} c_{2}^{2} M -\frac{1}{2}kc_1^2
\end{equation}

Eq. \eqref{eq:eq23} indicates that if there is no light stimulation, i.e. $M = 0$, there is still a negative energy term related to the membrane stiffness $k$. Disassembling Eq. \eqref{eq:eq23} into a combination of two items, we have the following equation:
\begin{equation}\label{eq:eq24}
	\varepsilon=\varepsilon_{s}-\varepsilon_{r}
\end{equation}

\textbf{\emph{Definition 1 (Repulsive Energy):}} \emph{During cell vibration, the energy used by the cell system to reject the stimulus is}
\begin{equation}\label{eq:eq25}
	\varepsilon_{r}=\frac{1}{2} k c_{1}^{2}
\end{equation}
where the minus sign indicates that the cell exhibits spontaneous resistance or repulsion during energy absorption and vibration. This is consistent with the medical observation that live cells are able to respond to external stimuli.

\textbf{\emph{Definition 2 (Single Stimulation Energy):}} \emph{When a single photon stimulates a retinal cell, the energy that cells can perceive by its membrane vibration behavior is:}
\begin{equation}\label{eq:eq5}
	\varepsilon_{s}=c_{1} c_{2} \sqrt{k M}-\frac{1}{2} c_{2}^{2} M
\end{equation}

Considering the vibration frequency $f$ of the cell membrane, the energy of the photon flow per unit cycle is:
\begin{equation}\label{eq:eq7}
	E=f \varepsilon_{s}=\frac{w_{0}}{2 \pi} \varepsilon_{s}
\end{equation}

Therefore, it is easy to deduce the expression of $E$.

\textbf{\emph{Definition 3 (Cycle Stimulation Energy):}} \emph{The energy response of retinal cells to photon flow stimulation during the unit cycle is:}
\begin{equation}\label{eq:eq19}
E=\frac{1}{2 \pi} c_{1} c_{2} k-\frac{1}{4 \pi} c_{2}^{2}\sqrt{k M}
\end{equation}

After {\color{blue}analyzing} the cell membrane vibration, we derive an explicit model consisting of Eqs. \eqref{eq:eq5} and \eqref{eq:eq19}, which {\color{blue}measure} part of the energy that {\color{blue}retinal cells transformed by} the vibration effect when stimulated by photons. Next, we will discuss how the proposed energy model works in the field of low-light image enhancement.

\subsection{Global Lightness Enhancement}
Assume that dim areas of the low-light image are marked $\Omega$, pixel $P \in \Omega$ and its value is $I$, $I\neq0$, the lightness mapping function is $\Psi(\cdot)$, then the mapping function for the constraint of the lightness enhancement process satisfies:
\begin{equation}\label{eq:eq20}
	\Psi(I) > I
\end{equation}

For Eqs. \eqref{eq:eq5} and \eqref{eq:eq19}, we rewrite them as:
\begin{equation}\label{eq:eq6}
	\varepsilon_{s}=\lambda \sqrt{I}+(1-\lambda) I
\end{equation}
\begin{equation}\label{eq:eq8}
	E=\frac{\lambda^{2}}{\sqrt{\lambda-1}}-\lambda \sqrt{(\lambda-1) I}
\end{equation}
where $I \in(0,1]$ and $\lambda \in(1,2]$. The condition for rewriting is $\varepsilon_{s}(I=0)=0, \varepsilon_{s}(I=1)=1, c_{1}=\frac{1}{2 \sqrt{2} \pi}, c_{2}=\sqrt{2 (\lambda-1)}, k=\frac{4\pi^{2}\lambda^{2}}{\lambda-1}, M=I$. The derivation of Eqs. \eqref{eq:eq6} and \eqref{eq:eq8} can be {\color{blue}found in} {\color{blue}\textbf{Appendix A}}. It is easy to know that Eq. \eqref{eq:eq6} satisfies the constraint, allowing it to perform lightness enhancement. The motivation for applying the energy model to the image field is that we assign a retinal cell with the same parameters at each pixel location in the image, and the pixel values are considered to be stimulus intensity signals ($M=I$). Therefore, the signal is transformed into a {\color{blue}nonlinear} energy signal. This is similar to the human visual system. In this paper, the parameters $c_{2}$ and $k$ are replaced by the parameter $\lambda$, which can be called the joint factor. It combines the roles of the parameters $c_{2}$ and $k$, as well as $\sqrt{I}$ and $I$.

In addition, the legacy gamma correction model\cite{b63} also satisfies the constraints. In the past, gamma correction models were often used to mimic the {\color{blue}nonlinear} working mechanism of human vision and display devices. For example, it is often used to correctly reproduce the luminance of CRT displays\cite{b60}. However, there is a distinct defect in the gamma correction model, that is, the image is easily fogged. Zhu et al.\cite{b61} used this defect feature to generate multiple exposures of a hazy image to obtain a {\color{blue}clear image}. Since $\varepsilon(\lambda,I)$ is less capable of mapping nonlinearity than the gamma correction model, $E(\lambda,I)$ has a curve property that decreases with {\color{blue}increasing} $I$. Therefore, a modified gamma correction model combining the energy model is proposed as a lightness enhancement mapping function. The proposed model is as follows:\begin{equation}\label{eq:eq10}
	C=\frac{I^{\Gamma} E(\lambda, I)}{I_{\max }^{\Gamma} E\left(\lambda, I_{\max }\right)}
\end{equation}
{\color{blue}, where $C$ is} the output of the modified model, and $\Gamma$ is the gamma intensity, which is used to expand the {\color{blue}nonlinear} mapping space. We have improved the above traditional gamma correction models by fusing the two models. $E$ is inversely proportional to $I$. When $I$ is larger, the value of $E$ is smaller. $E(\lambda,I)$ can be thought of as an adjustment function where the areas with low pixel values are assigned a high product factor while the areas with high pixel values have small coefficients, and then simple normalization measures are used to ensure that the mapping domain falls in the range of $[0,1]$. This approach readjusts the mapping space of the gamma correction function and compresses the increase in the high value range. 

Such a function can be constructed based on empirical evidence, for example, $Hx^{(Jy+K)}$. However, appropriate system parameters are very difficult to {\color{blue}determine} in such a way.
Fig. \ref{fig9} shows the advantages of the proposed model, which proves that this technique is effective.

Further studies have found a new statistical property associated with the proposed gamma model. Detailed descriptions are as follows:

\textbf{\emph{Statistical Property:}} \emph{For the V-channel in the HSV space, we assume that the average lightness of the {\color{blue}nonzero} elements in the input image is $I_{vm}$, the average lightness of the {\color{blue}nonzero} elements in the output image is $C_{vm}$, and the lightness difference is $\Delta V= C_{vm} - I_{vm}$. Given $\lambda$, there is a transformation between the lightness difference and the gamma intensity, and the mathematical model is represented as:}
\begin{equation}\label{eq:eq11}
	\Delta V=c+\frac{1}{a \Gamma+b}
\end{equation}

\begin{equation}\label{eq:eq35}
	C_{v m}=\frac{1}{N-N_{0}} \sum C_{L}(x, y)
\end{equation}
\begin{equation}\label{eq:eq36}
	I_{v m}=\frac{1}{N-N_{0}} \sum I_{L}(x, y)
\end{equation}
where the curve parameters (a, b and c) are different for different input images. $C_{L}$ is the V-channel image of the output image $C$ in HSV space, and $I_{L}$ is the V-channel image of the input image in HSV space. $C_{L}(x, y)$ and $I_{L}(x, y)$ are the lightness values at coordinates $(x, y)$. $N$ is the total number of pixels in image $I_{L}$, and $N_{0}$ is the number of pixels with a lightness value of 0.
Given the lightness difference $\Delta V$ between the output image and the input image, the corresponding gamma intensity can be recovered according to Eq. \eqref{eq:eq11}, which adjusts the lightness of the low-light image. 

Because different images have different curve parameters, if different images need to achieve the same lightness difference, the gamma intensity needs to be adjusted for each image. Based on statistical properties, this adjustment has been automated to avoid tedious manual tuning.
This is very similar to the dynamic lightness regulation by retinal cells. 

It should be noted that Eq. \eqref{eq:eq11} is an empirical formula based on statistical experiments. Some statistical evidence is disclosed in Experiment B: statistical cases.

To obtain the $\Gamma - \Delta V$ curve of the current input image, we used a perception strategy.

\textbf{\emph{Perception Strategy:}} \emph{For the V-channel in the HSV space, given $\lambda$, we preset a series of gamma intensities, such as sequence $\Gamma_{s}=\left[\Gamma_{1}, \Gamma_{2}, \ldots, \Gamma_{n}\right]$. For each gamma intensity, the input image is calculated by Eq. \eqref{eq:eq10} to obtain n copies of the output image. Then, the average lightness $C_{vm}$ of {\color{blue}nonzero} elements in each output image copy is calculated, and a series of lightness differences $\Delta V_{s}=\left[\Delta V_{1}, \Delta V_{2}, \ldots, \Delta V_{n}\right]$ can be obtained. By using sequence $\Gamma_{s}$, sequence $\Delta V_{s}$ and considering efficiency, we tend to use a nonlinear least square method called the trust region reflection algorithm\cite{b44, b45} to render the parameters in the curve model.}

\begin{figure}[bp]
	\centerline{\includegraphics[width=\columnwidth]{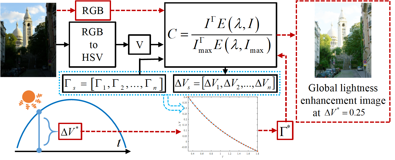}}
	\caption{Global lightness enhancement scheme of the low-light image. Black solid lines represent the process of calculating $\Delta V_{s}$, blue dashed lines represent the curve capture phase, and red dashed lines represent the global lightness enhancement phase.}
	\label{fig15}
\end{figure}

\begin{figure}[h]
	\centerline{\includegraphics[width=\columnwidth]{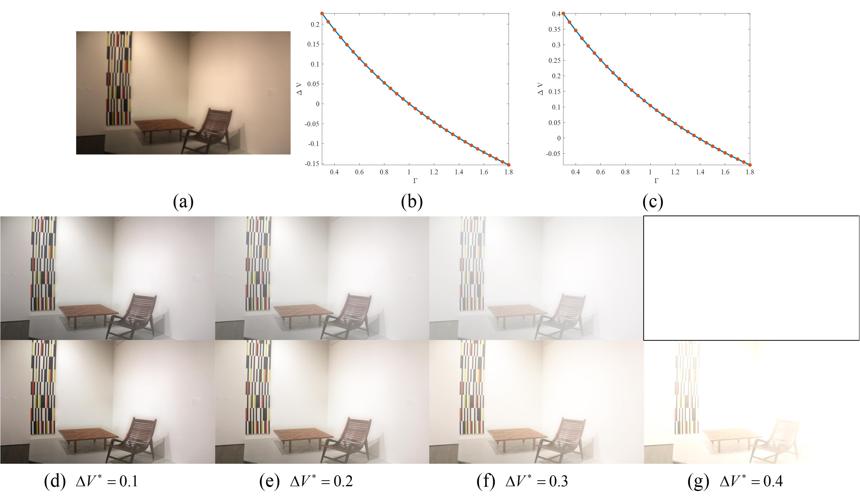}}
	\caption{The difference between the two models. (a) Low-light image from an HDR dataset. (b) Statistical property curve of the traditional model. (c) Statistical property curve of the proposed model. Note that the {\color{blue}solid blue line} represents the true value, and the red dot represents the fitting value. (d $\sim$ g) A series of global lightness enhancement images generated under $\Delta V^{*}$ constraints, with the upper row representing the traditional model and the lower row representing the proposed model.}
	\label{fig9}
\end{figure}

Based on the above analysis, we summarize the entire framework for global lightness enhancement. Fig. \ref{fig15} graphically depicts the entire scheme.

\textbf{\emph{Global Lightness Enhancement Scheme:}} \emph{For the input low-light image, given $\lambda$ and gamma intensity sequence $\Gamma_{s}=\left[\Gamma_{1}, \Gamma_{2}, \ldots, \Gamma_{n}\right]$, we assume that the lightness difference between the output and the input images is $\Delta V^{*}$. The scheme consists of two {\color{blue}phases. The first phase} is to obtain the curve parameters, and the second phase is to enhance the global lightness.}
	
\emph{In the first phase, we first obtain the V-channel of the low light image and then use the perception strategy to obtain the lightness difference sequence $\Delta V_{s}$ and the three parameter values in the statistical property curve. After obtaining the parameters of the statistical property curve, we can see how the lightness of the global light changes with the gamma intensity.}

\emph{In the second phase, to obtain a global enhanced image $C$ with a lightness difference of $\Delta V^{*}$, we first use the statistical property curve of the first phase to obtain the gamma intensity $\Gamma^{*}$ at $\Delta V=\Delta V^{*}$. In this way, a special mapping curve of Eq. \eqref{eq:eq10} at $\Gamma=\Gamma^{*}$ is obtained. Finally, this special mapping curve is used to calculate each channel of the low light image in the RGB space to obtain the global enhanced image $C$.}

Fig. \ref{fig9} shows a series of global lightness enhancement images obtained with the lightness difference $\Delta V^{*}$ constraint and the statistical property curves of the two models. Traditional gamma correction requires a smaller gamma intensity to achieve the same lightness enhancement. This exacerbates the atomization effect, resulting in a significant decrease in image saturation, and the local features begin to become indistinguishable or even completely lost. After we have applied the proposed energy model, image degradation is reduced, and many features are retained with significant brightness enhancement. The statistical property curve of the proposed model also shows that large lightness enhancement can be achieved without requiring a small gamma intensity.

\begin{figure}[t]
	\centerline{\includegraphics[width=\columnwidth]{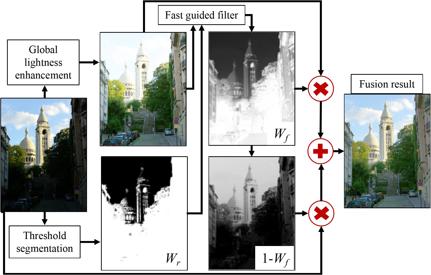}}
	\caption{The whole process of lightness enhancement for a single low-light image and the results of each phase of the local fusion strategy}
	\label{fig1}
\end{figure}

We proposed a new global lightness enhancement model to de-expose the low-light areas and improve the drawbacks of traditional gamma correction models. At the same time, we noticed that the proposed global model could not handle high-light areas. Next, we propose a local fusion strategy based on fast-guided filtering to improve the local details of the global image.

\subsection{Local Fusion Strategy}
In this subsection, we propose a simple, fast and effective local fusion strategy to further optimize the global image. The proposed strategy consists of three phases: segmentation, filtering and fusion. First, we use segmentation to obtain a mask image (also known as a rough weight map), which depicts the general distribution of light and dark areas in the input image. Then, we use a fast guided filter to obtain fine fusion weights, which can restore the local spatial manifold of the mask image and transform {\color{blue}a hard boundary into a soft boundary}. Finally, the defective features in global lightness enhancement images are replaced by mask fusion. {\color{blue}A detailed} description is as follows:


(1) \textbf{Segmentation}: For the V-channel of the input image, {\color{blue}downsampling} is performed first, and then the {\color{blue}downsampled} image is segmented using a hard threshold to obtain a rough weight map. The segmentation rule we used is:
\begin{equation}\label{eq:eq12}
W_{r}(x, y)=\left\{\begin{array}{cc}
1, & if I_{v}(x, y)<T \\
0, & else
\end{array}\right.
\end{equation}
where $W_{r}$ is the rough weight map, {\color{blue}$I_{v}$ is the V-channel downsampled result of the input low-light image, and the down sampling rate is $R$, which is used to reduce the amount of data in the segmentation process. $T$ is the segmentation threshold.} $W_{r}(x, y)$ and $I_{v}(x, y)$ represent the values of $W_{r}$ and $I_{v}$ at coordinates $(x, y)$, respectively. Since the effective range of the luminance space is [0,1], we set $T = 0.5$ to divide the whole space into two subspaces, light and dark. Areas with values greater than $T$ are {\color{blue}considered} light areas, while a coarse weight of 0 is assigned. {\color{blue}When the threshold $T$ is set to a different value (not 0.5), the direct consequence is that the scale range of light and dark areas is no longer equal, which results in a biased information perception. Therefore, threshold $T = 0.5$ is a fair solution.}

(2) \textbf{Filtering}: The V-channel downsampled result of the global lightness enhancement image is the guidance image. The coarse weight graph is filtered by a fast guided filter, and the filter results are upsampled to obtain a fine weight map. The filtering process can be simplified as:
\begin{equation}\label{eq:eq13}
W_{f}=G_{f}\left(C_{v}, W_{r}, w, r, \eta\right)
\end{equation}
Here, $G_{f}(\bullet)$ represents the fast guided filtering process. $C_{v}$ acts as a guidance image to correct the local spatial manifold of the rough weight map $W_{r}$. $w$ is the radius of the square filter's window, $r$ is the down sampling rate of the fast guided filter and $\eta$ is a custom regularization constant, usually $\eta=0.04$. The fast guided filter is a {\color{blue}classic and rapid computation} filter\cite{b57,b58}. Finally, the fine weight map is obtained after an {\color{blue}upsampling} operation with a sampling rate $R$ is performed on the filtered results $W_{f}$. 

If there is no region larger than the segmentation threshold in the image, the image is regarded as a dark partition globally, but the spatial local manifold can still be found, and the appropriate fusion weight can be obtained by the fast guided filter. The image $X_{1}$ shown in Fig. \ref{fig7} is an example of this case. Although there is no light partition exceeding the threshold, our system can still work normally, and the resulting image will not have image quality problems such as unsmoothed artifacts.

(3) \textbf{Fusion}: The output of the global lightness enhancement scheme is fused with the input image to form a harmonious and natural image. The local fusion formula is:
\begin{equation}\label{eq:eq1}
\begin{aligned}
	F_{i}=&W_{f} \cdot C_{i}+\left(1-W_{f}\right) \cdot X_{i} \\
	=     &W_{f} \cdot \frac{X_{i}^{\Gamma^{*}} \cdot E\left(\lambda, 	  X_{i}\right)}{X_{i\max }^{\Gamma^{*}} \cdot E\left(\lambda, X_{i\max }\right)}+\left(1-W_{f}\right) \cdot X_{i}
\end{aligned}
\end{equation}
where $X$ is the input image, $F$ is the output image, and $i$ represents the $ith$ channel in the RGB space. $\cdot$ is the {\color{blue}elementwise} product, and $X_{i}^{\Gamma^{*}}$ represents the result of the gamma transformation of $X_{i}$ when the gamma intensity is $\Gamma^{*}$. Fig. \ref{fig1} is a typical example of how a local fusion strategy works with the proposed modified model to achieve lightness enhancement. The sky of the input image $X$ in Fig. \ref{fig1} seems clean. After enhancing lightness, details appear in the dark, but the sky details are lost. By using the local fusion strategy, the sky features in the input image are successfully fused, resulting in clear and well-balanced images.

De-masking is a direct and effective remedy solution, which has a significant advantage over the iterative solutions {\color{blue}on with} low computational cost. Subsequent experiments have confirmed this analysis.

\subsection{Discussion}
\emph{Why replace the parameters in the energy model with parameter $\lambda$?}
The parameters $c_{1}$, $c_{2}$, and $k$ are used to describe the movement and intrinsic properties of real retinal cells. The values of $c_{1}$, $c_{2}$, and $k$ need to be determined in physical measurement experiments. For different types of cells, we assume that the values of these parameters are different. Therefore, to use the proposed energy model, we used the {\color{blue}parameter} $\lambda$ to rewrite the equation.

\emph{How large should the filter window be?}
The window radius $w$ is related to the amount of local detail in the weight map $W_{f}$. As the window size increases, the local feature information perceived by the window increases, which provides guidance for correcting local spatial manifolds in the rough weight maps $W_{r}$. To obtain a fine weighted map with rich local details, through experimental evaluation, the window radius $w$ we select is
\begin{equation}\label{eq:eq14}
w=\lfloor\frac{\min (row, col)-1}{2}\rfloor
\end{equation}
$\lfloor\bullet\rfloor $ is a downward rounding operation to avoid the window size exceeding the limit. $row$ represents the number of rows of image $C_{v}$, and $col$ represents the number of columns of image $C_{v}$.

\begin{figure}[ht]
	\centerline{\includegraphics[width=\columnwidth]{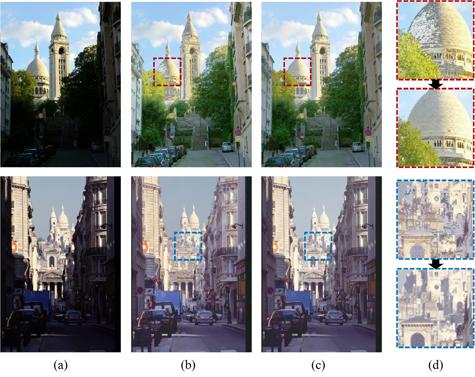}}
	\caption{Fusion results of different weight maps. (a) Low-light image with uneven illumination from an ExDark dataset. (b) Defective fusion results due to rough weights. (c) Harmonious images obtained using fine weights. (d) {\color{blue}Detailed} display of the local patch.}
	\label{fig10}
\end{figure}

\emph{How does a coarse or fine weight map affect the final fusion result?} 
Fig. \ref{fig10} shows the fusion results using coarse and fine weight maps. Rough weight maps produce disharmonious fusion results because local manifold characteristics are not taken into account. With the help of the fast guided filtering method, an estimation solution of the local spatial manifold structure is found, which successfully solves the problem caused by rough fusion and makes the fused image appear natural and harmonious.


\section{Experimental Work}
\label{sec:exp}
In this section, we first examine the impacts of the involved parameters. 
Next, we present statistical evidence supporting Eq. \eqref{eq:eq11}. Afterwards, we compare our method with several {\color{blue}state-of-the-art} methods from the perspective of subjective qualitative and objective quantitative evaluation. Finally, we give the time consumption of different enhancement methods. The comparison methods include HE\cite{b7}, NPEA\cite{b46}, SRIE\cite{b47}, LIME\cite{b2}, AIEMC\cite{b14}, ASTF\cite{b48}, MF\cite{b51}, EFF\cite{b52}, and KinD\cite{b33}. A test platform based on a computer (AMD Ryzen 5 2400G @3.6 GHz, 16 GB RAM and {\color{blue}Windows} 10 operating system) is established. The software was developed with MATLAB 2018b. To compare the methods, we use the MATLAB code published by the authors. The KinD {\color{blue}PyTorch} code is provided by the author, and the NVIDIA GeForce RTX 3060 (6 GB) graphics card is used to speed up the calculation. The images used in the experiment come from publicly accessible datasets. The majority of our experimental images are from four databases, SIDD\cite{b35,b37}, the ESPL-LIVE HDR Image Quality Database (HDR)\cite{b38,b39}, ExDark\cite{b42}, NASA\cite{b59}, LOL\cite{b71} and a synthetic dataset\cite{b71}.

\subsection{Parameters' impact}
The parameters to be configured in the global lightness enhancement stream are $\lambda$, sequence $\Gamma_{s}$, and $\Delta V^{*}$. The parameters to be discussed in the local fusion strategy are $R$, $r$ and $w$. Through experimental analysis, a set of appropriate parameters are configured as $\lambda=2$, $\Gamma_{s}=[0.3,0.8,1.3,1.8]$, $\Delta V^{*}=0.25$, $R=2$, and $r=10$. The window radius $w$ is shown in \eqref{eq:eq14}. This set of parameters is used for subsequent qualitative evaluation {\color{blue}experiments} and objective quantitative {\color{blue}examinations}. Next, we will discuss their impact separately.

\begin{figure}[h]
	\centerline{\includegraphics[width=\columnwidth]{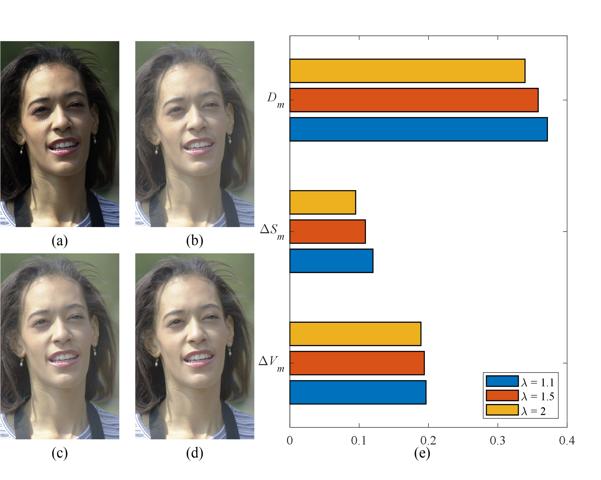}}
	\caption{{\color{blue}The influence of parameter $\lambda$ on fusion results.} (a) Input image with uneven illumination from a NASA dataset. (b $\sim$ d) Image results when $\lambda$ is set to 1.1, 1.5, and 2. (e) Horizontal bar chart of three statistical states.}
	\label{fig2}
\end{figure}

The parameter $\lambda$ is configured as 1.1, 1.5 and 2. The other parameters are unchanged. The range of parameter $\lambda$ is $(1,2]$. Because there are infinite parameter choices in a finite space, to demonstrate the system performance under different parameter values, we only show three possible cases, i.e., $\lambda$ = 1.1, 1.5 and 2. The results are shown in Fig. \ref{fig2}. The visual performance of the fusion results is satisfactory, and the difference is very small, except that the fusion results are gray when $\lambda=1.1$. A closer examination shows that the colors are more vivid when $\lambda=2$. More detailed numerical results are shown in Fig. \ref{fig2}(e). We use three statistical states, namely, the mean difference of lightness $\Delta V_{m}$, mean difference of saturation $\Delta S_{m}$ and difference of lightness and saturation $D_{m}$, to analyze the effect of parameter $\lambda$. The three statistical states are calculated as follows.
\begin{equation}\label{eq:eq18}
\left\{\begin{array}{c}
\Delta V_{m}=F_{v m}-I_{v m}\\
\Delta S_{m}=I_{s m}-F_{s m}\\
D_{m}=F_{v m}-F_{s m}
\end{array}\right.
\end{equation}
where $F_{v m}$ and $F_{s m}$ are the average lightness and saturation of the fused image, respectively. $I_{s m}$ is the average saturation of a low-light image. Fig. \ref{fig2}(e) shows that $\Delta V_{m}$ has decreased from 0.25 to approximately 0.2 due to the low illumination of the low-light image. Because the difference between the three $\lambda$ is small, this means that $\lambda$ has a slight effect on $\Delta V_{m}$. The color attenuation prior proposed by Zhu\cite{b62} indicates that there is a saturation decrease, and $D_{m}$ increases during the transition from a clear image to a hazy image. Therefore, $\Delta S_{m}$ and $D_{m}$ of a clear and vivid fusion image are at a low level. Since $\Delta S_{m}$ and $D_{m}$ decrease significantly when $\lambda=2$, we choose $\lambda=2$ based on the above analysis.

\begin{figure}[!t]
	\centerline{\includegraphics[width=\columnwidth]{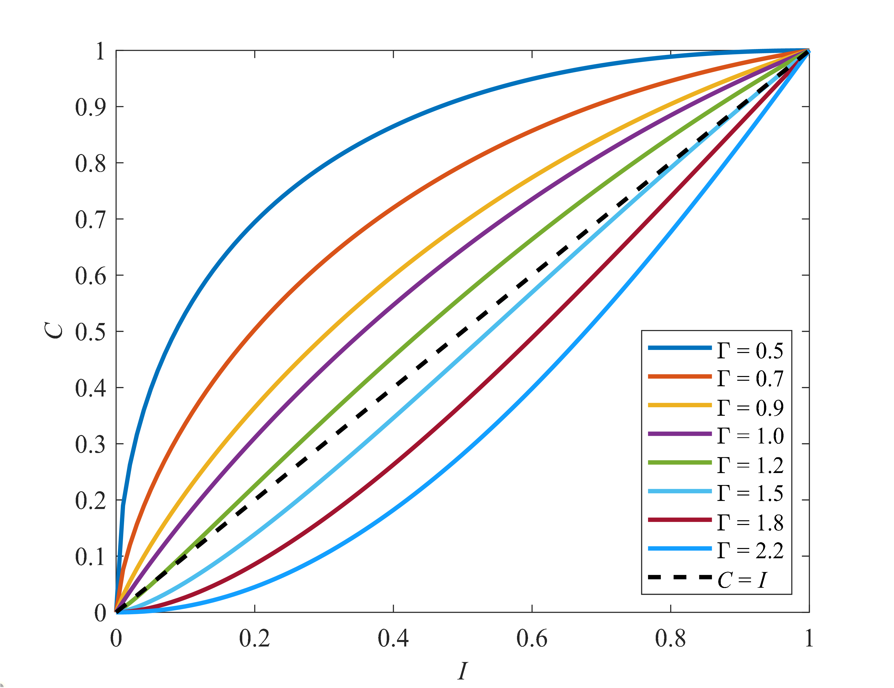}}
	\caption{$I-C$ curves at different gamma intensities $\Gamma$.}
	\label{fig3}
\end{figure}

{\color{blue}The gamma} sequence $\Gamma_{s}$ is a reference element of curve fitting. It represents a combination of gamma distributions used to obtain the three parameters of Eq. \eqref{eq:eq11}. To solve the parameters of the nonlinear model shown in \eqref{eq:eq11}, at least four reference coordinate points are required. We suggest that the range of preselected points for $\Gamma_{s}$ be [0.3, 2.2]. For Eq. \eqref{eq:eq10}, if $\lambda = 2$ and the maximum value is obtained at $I = 1$, the variation of $C$ with the gamma intensity can be obtained as shown in Fig. \ref{fig3}. 

Since the combinatorial number of gamma sequences is $C_{n}^{4}(n \rightarrow \infty)$, we only discuss and compare the following four gamma sequences for analysis purposes.

$S_{1}: \Gamma_{s}=[0.3,0.8,1.3,1.8]$; $S_{2}: \Gamma_{s}=[0.8,0.9,1.1,1.2]$;

$S_{3}: \Gamma_{s}=[0.3,0.5,0.6,0.8]$; $S_{4}: \Gamma_{s}=[1.3,1.5,1.6,1.8]$.

\begin{figure}[t]
	\centerline{\includegraphics[width=\columnwidth]{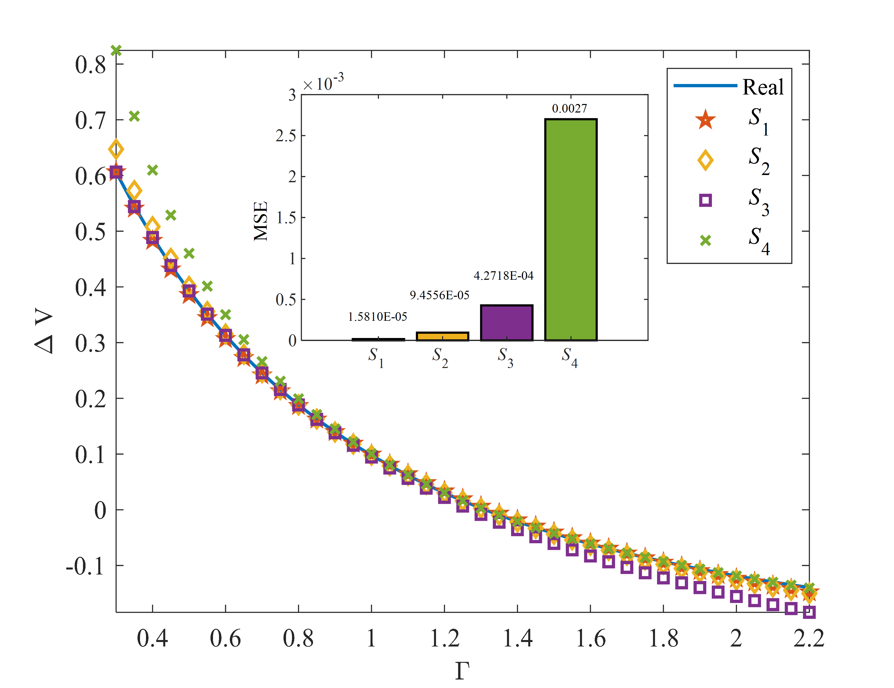}}
	\caption{Curve fitting and error of different gamma sequences.}
	\label{fig4}
\end{figure}

We continue to use Fig. \ref{fig2}(a) in the experiments. Fig. \ref{fig4} shows the fitting and the mean square error (MSE) of the four sequences. In these four cases, the sequence $S_{1}$ has the lowest fitting error, and the sequence $S_{4}$ has the worst fit. Therefore, we recommend the gamma sequence $S_{1}$.

\begin{figure}[h]
	\centerline{\includegraphics[width=\columnwidth]{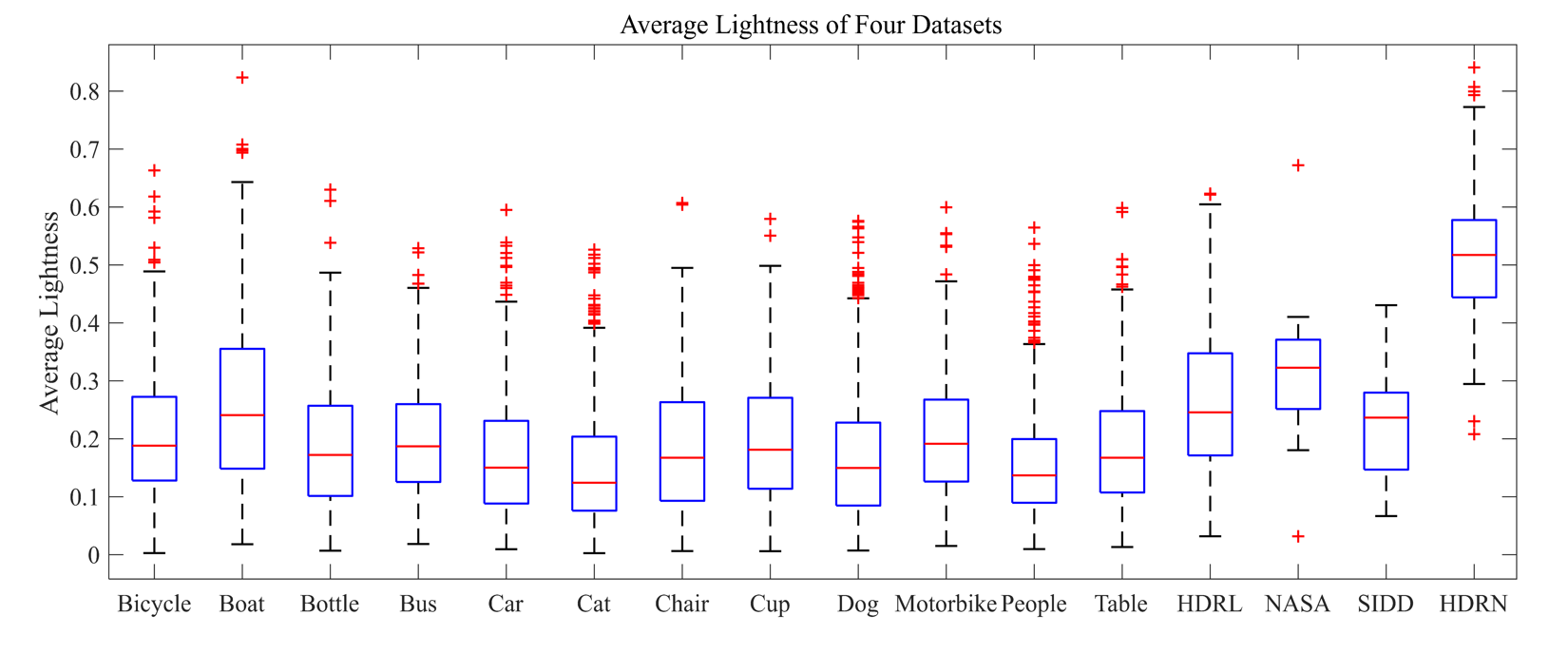}}
	\caption{Average lightness box chart for the four datasets. HDRL and HDRN represent a low-light image and normal light image in the HDR dataset, respectively. Each box is the 25th to 75th percentile of the sample. The red line is the sample median; an outlier appears as a red + sign. The black lines above and below the box are the upper and lower bounds of the sample.}
	\label{fig11}
\end{figure}

The lightness difference $\Delta V^{*}$ is an important parameter. Because $C_{vm}=I_{vm}+\Delta V^{*}$, the lightness of the global lightness enhancement image changes with $\Delta V^{*}$ and ultimately affects the lightness of the fused image. For low-light image enhancement, $\Delta V^{*}$ must be greater than 0. To select an appropriate lightness difference, we calculated the average lightness distribution of the four databases, and the results are shown in Fig. \ref{fig11}. Details are ExDark: 7363 images; HDR: 250 pairs of images (including low-light and normal); NASA: 22 images; SIDD: 74 images. The Exdark dataset contains 12 categories of images. These categories are bicycle, boat, bottle, bus, car, cat, chair, cup, dog, motorcycle, people and table. The numbers of these categories are 652, 679, 547, 527, 638, 735, 648, 519, 801, 503, 609 and 505. 

The statistical results show that the average lightness of most of the low-light images is less than 0.4, the median is {\color{blue}approximately} 0.2, and the average upper and lower bounds are 0.0623$\sim$0.4724. Most of the normal illuminated images fall between 0.4441 and 0.5776, and the upper and lower bounds are 0.29454$\sim$0.77257. It is recommended that $\Delta V^{*}$ be set between [0.2, 0.3]. This conclusion is consistent with the observation in Fig. \ref{fig9}; that is, the lightness enhancement of the image is insufficient when $\Delta V^{*}$ is less than 0.2, and it may be too bright when $\Delta V^{*}$ is greater than 0.3. In addition, it is assumed that for different images, $\Delta V^{*}$ conforms to a mathematical distribution within the range [0.2, 0.3], such as a normal distribution. Therefore, based on the above analysis, we recommend $\Delta V^{*}=0.25$.

\begin{figure}[!t]
	\centerline{\includegraphics[width=\columnwidth]{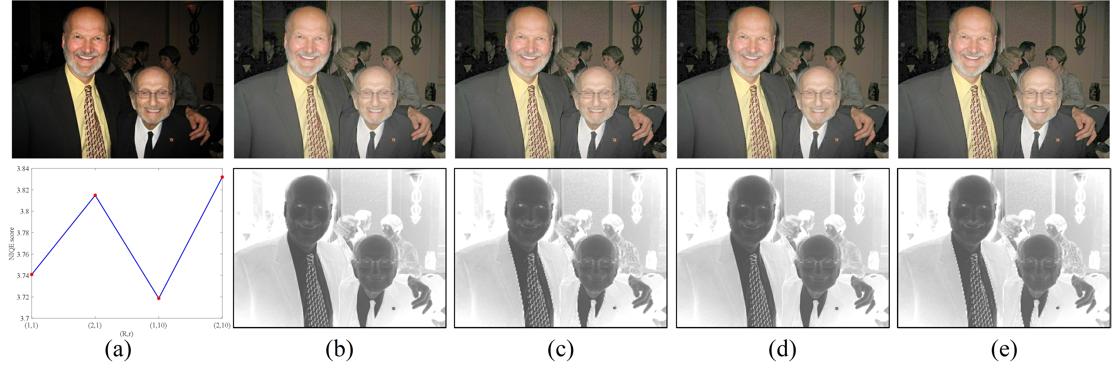}}
	\caption{Weight maps and NIQE scores for different down sampling rates. (a) Top: Input image with uneven illumination from the Exdark dataset. Bottom: NIQE scores for different down sampling rates. (b)$\sim$(e) The result at $(R, r) =$ (1,1), (2,1), (1,10) and (2,10), respectively. From top to bottom are the fusion results and weight maps. The black border is artificially added to highlight the boundary. The NIQE scores of the four combinations differ slightly, with a maximum deviation of {\color{blue}approximately} 0.11. The NIQE score shows that the influence of {\color{blue}the downsampling} rate $R$ on the images is more significant than that of {\color{blue}the downsampling} rate $r$.}
	\label{fig5}
\end{figure}

Downsampling rates $R$ and $r$ are set up to speed up the algorithm. To speed up the acquisition of rough weight maps, we downsample the V-channel of the input image. On this basis, a fast guided filter is used to further accelerate the calculation. The downsampling rate $r$ of the fast-guide filter does not significantly change the filtering result. {\color{blue}Therefore}, we set $r = 10$. Fig. \ref{fig5} shows the result of the weight map generated by different combinations of downsampling rates $R$ and $r$. Obviously, $R$ significantly affects the local details of the fine weight map. When $R = 2$, there is a small loss of detail in the weight map, and an increase in the NIQE score also means a decrease in image quality. {\color{blue}However}, the difference between the fusion results of the four combinations is very small, and it is difficult for human eyes to detect the difference. Therefore, we recommend setting $R=2$ to speed up the algorithm calculation while allowing the weight map to have a small loss of detail. 

\begin{figure}[t]
	\centerline{\includegraphics[width=\columnwidth]{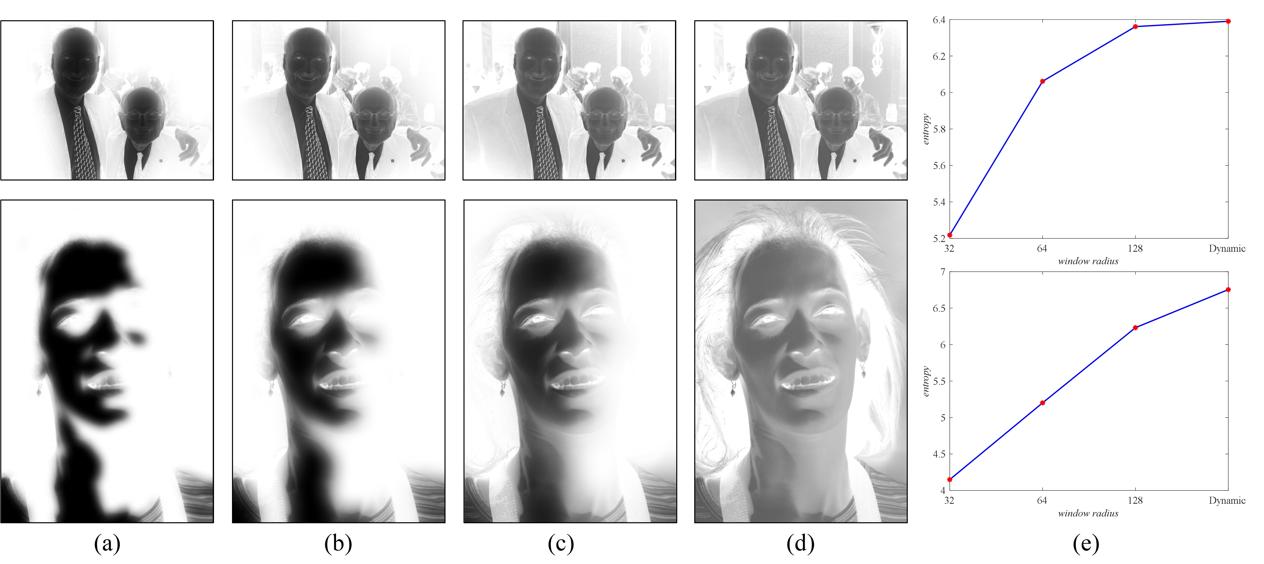}}
	\caption{Weight maps and entropy for different window radii. The dimensions from top to bottom are $375 \times 500 \times 3$ and $2000 \times 1312 \times 3$, respectively. (a)$\sim$(c) Weight maps are generated with 32, 64, and 128 filter radii, respectively. (d) The weight map is generated by using a dynamic window radius. (e) Entropy of four window radii. Top: Image "group photo". Bottom: Image "smiling woman". The entropy metric\cite{b73} quantitatively describes the impact of different window sizes on the weight map. The black border is artificially added to highlight the boundary.}
	\label{fig6}
\end{figure}

\begin{table*}[t]
	\caption{Statistical results of curve fitting error (MSE) for 12 categories of images in Exdark datasets(7363 images). Above double horizontal line: the proposed model. Below double horizontal line: gamma correction model.}
	\label{table1}
	\setlength{\tabcolsep}{3pt}
	\begin{tabular}{p{40pt}<{\centering}p{30pt}<{\centering}p{30pt}<{\centering}p{30pt}<{\centering}p{30pt}<{\centering}p{30pt}<{\centering}p{30pt}<{\centering}p{30pt}<{\centering}p{30pt}<{\centering}p{30pt}<{\centering}p{30pt}<{\centering}p{30pt}<{\centering}p{30pt}<{\centering}p{35pt}<{\centering}}
		\hline Item & Bicycle & Boat & Bottle & Bus & Car & Cat & Chair & Cup & Dog & Motorbike & People & Table & Exdark \\
		\hline max($\times$$10^{-4}$) & 4.6261 & 4.2539 & 6.4014 & 4.8703 & 5.6659 & 6.8252 & 5.2790 & 5.6211 & 6.1084 & 4.8257 & 5.5835 & 6.0037 & 6.8252\\
		\hline min($\times$$10^{-8}$) & 4.3907 & 1.1844 & 3.0893 & 7.4585 & 6.3653 & 3.1414 & 0.5538 & 2.3780 & 4.5391 & 0.4970 & 5.9148 & 2.5236 & 0.4970\\
		\hline mean($\times$$10^{-5}$)  & 8.7455 & 6.5737 & 9.6162 & 8.8055 & 11.8130 & 11.0900 & 9.5539 & 8.2799 & 10.3410 & 8.7760 & 11.8260 & 9.4859 & 9.5756\\
		\hline \hline max($\times$$10^{-4}$) & 1.3183 & 1.2009 & 1.8465 & 1.3704 & 1.6543 & 1.9085 & 1.5773 & 1.6557 & 1.7769 & 1.4112 & 1.6621 & 1.7367 & 1.9085\\
		\hline min($\times$$10^{-9}$) & 8.2443 & 15.5261 & 0.6561 & 15.9902 & 6.4102 & 18.0143 & 1.4632 & 0.2506 & 5.6929 & 46.0172 & 3.8311 & 9.3132 & 0.2506\\
		\hline mean($\times$$10^{-5}$)  & 2.6555 & 1.9880 & 2.8230 & 2.5839 & 3.4671 & 3.6489 & 2.9047 & 2.4707 & 3.3271 & 2.6558 & 3.5561 & 2.8684 & 2.9124\\
		\hline	
	\end{tabular}
	\label{tab1}
\end{table*}

For the window radius $w$, we do not recommend a fixed window size but use a dynamic window size for different images. This is because the spatial distribution range of image features changes with the size of the image, which makes the small filter window in the small-scale image unable to adapt to the large features in the large-scale image. Therefore, due to the incomplete collection of local spatial manifold information, some spatial manifold information is lost, and finally, the local manifold of the weight map is not sound. Considering that the larger the window is, the more characteristic information it will collect, we use the dynamic window radius shown in Eq. \eqref{eq:eq14}. Fig. \ref{fig6} shows the weight map generated by two different sizes of images at different window sizes. It is easy to know that the dynamic window radii are 187 and 655, respectively. When the window radius is 32, only a few local spatial manifold structures are found. With the increase in the window radius, the local space manifold becomes stable gradually and reaches the final state at the dynamic window radius. In addition, increasing entropy also means that more manifold structural information is found. The fused result has a flaw similar to Fig. \ref{fig10} when the window is small, so there is no additional display of the result image to save space. Similarly, the fusion results of the dynamic window are shown in Fig. \ref{fig2}(d) and Fig. \ref{fig5}(e).

\subsection{Statistical Cases}
The Exdark dataset is used to verify the validity of the statistical property and to illustrate the rationality of Eq. \eqref{eq:eq11}. Specifically, we examined a total of 7363 images from 12 categories of the Exdark dataset. For any image, the sequence $S_{1}$ is used to obtain the curve parameters. Then, we calculated the MSE of the gamma interval [0.3,2.2] with a step size of 0.05. The maximum, minimum and mean values in the MSE for each category and for the entire dataset are recorded in Table \ref{tab1}. It should be noted that the MSE metric is only used to verify the rationality of Eq. \eqref{eq:eq11}. For the proposed model \eqref{eq:eq10} and the gamma correction model, the fitting error on the whole dataset is very small, which indicates that for a natural low-light image, using the proposed model or gamma correction, a series of results can form a curve as described by the statistical property. It also shows that the mathematical expression of Eq. \eqref{eq:eq11} is reasonable.
The proposed method has a slightly larger MSE value than the gamma correction model, with MSE values ranging from $10^{-4}$ to $10^{-8}$. This result proves the validity of Eq. \eqref{eq:eq11} with a low MSE indicator score.
However, the statistical property is not always valid for a local block of the image. If the element value of a local block is 1, it remains the same throughout the process. Therefore, if this local block is treated as an image separately, it does not have statistical properties.

\subsection{Subjective and qualitative evaluations}
\begin{figure*}[!t]
	\centerline{\includegraphics
		[width=1.0\textwidth]{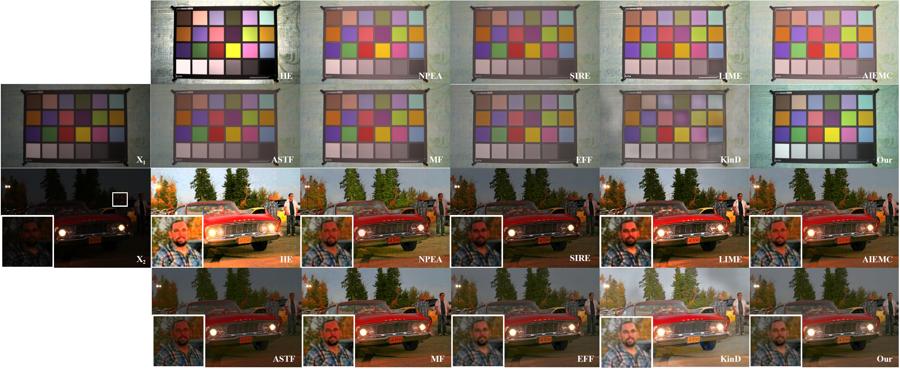}}
	\caption{Comparison results for images. {\color{blue}Images} X$_{1}$ and X$_{2}$ are taken from the SIDD and Exdark datasets, respectively. In the lower left corner of the image is a local detail image with artificially added white boxes to distinguish the differences.}
	\label{fig7}
\end{figure*}

\begin{figure*}[!t]
	\centerline{\includegraphics
		[width=1.0\textwidth]{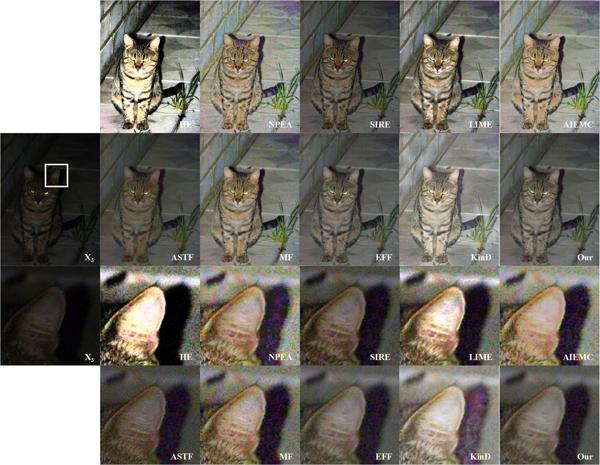}}
	\caption{Comparison results for noisy image. Image X$_{3}$ is taken from the Exdark dataset. The third and fourth rows are local feature images of the cat's ear.}
	\label{fig12}
\end{figure*}

\begin{figure*}[!t]
	\centerline{\includegraphics
		[width=0.88\textwidth]{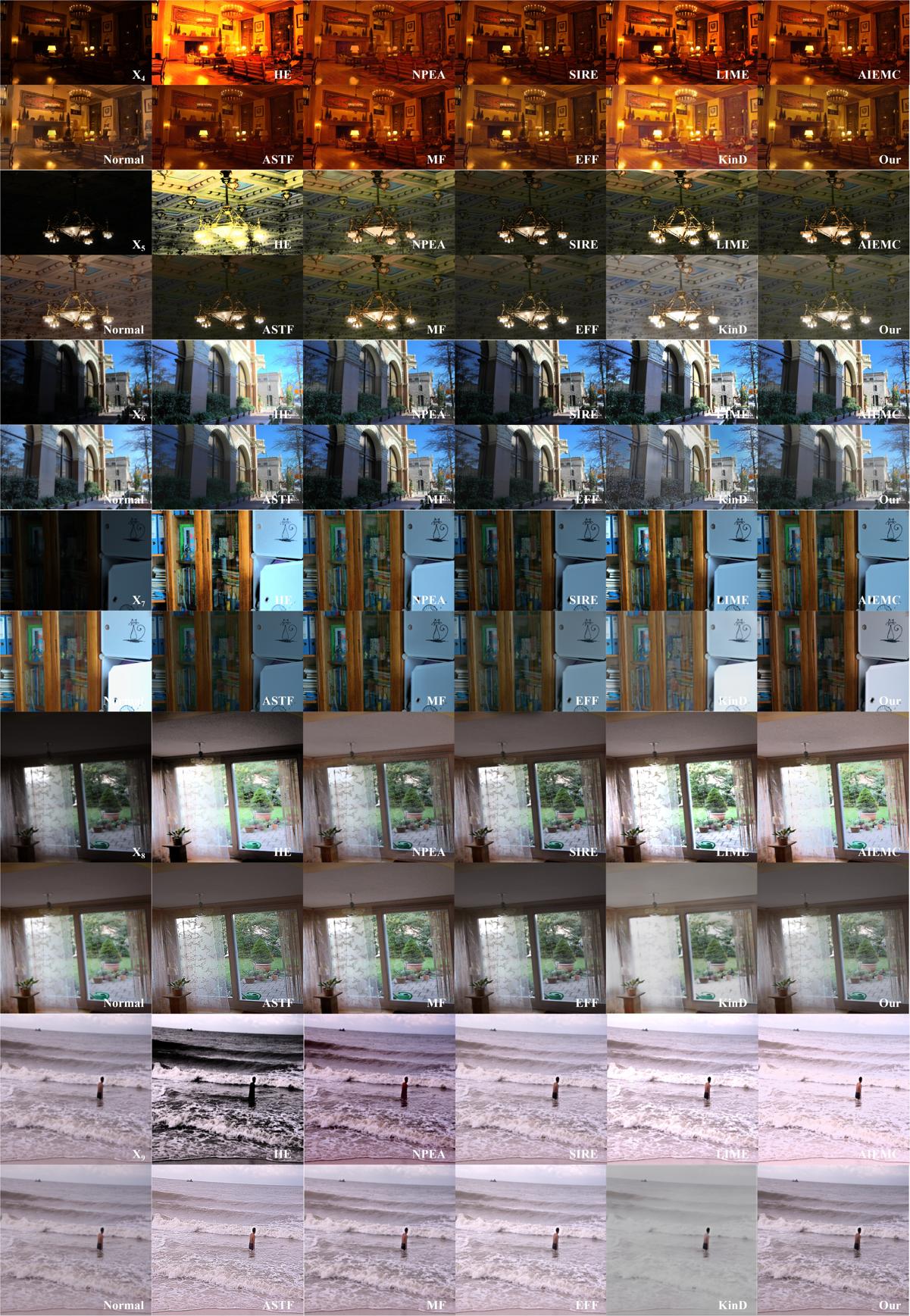}}
	\caption{Comparison results for image examples. {\color{blue}Images} X$_{4}$, X$_{5}$, X$_{6}$ and normal images are taken from the HDR dataset. Image X$_{7}$ is taken from the LOL dataset. {\color{blue}Images} X$_{8}$ and X$_{9}$ are taken from the synthetic dataset. Normal represents the image under normal illumination.}
	\label{fig8}
\end{figure*}

We selected images from the SIDD, Exdark, HDR, LOL and synthetic datasets as {\color{blue}references}. Figs. \ref{fig7}, \ref{fig12} and \ref{fig8} show the results of the nine algorithms. Image X$_{1}$ is a color correction board in a low-light state, which can be used to compare the color recovery capabilities of different algorithms. Our method yields clear and appropriate colors, and other methods result in a distorted appearance, i.e. white squares are significantly reddish or gray. 

This indicates that the results of other methods are biased due to the compression or expansion of the {\color{blue}grayscale}. For image X$_{1}$, the image size is 5312$\times$2988$\times$3. The KinD method cannot deliver the calculation due to the limitation of the video memory, so we downsample the image to 960$\times$540$\times$3. However, the result of the KinD method does not obtain a significant lightness improvement, and uneven color patches appear.
For image X$_{2}$, our results are very similar to EFF. Local facial details and colors show that the results of our method and EFF are more in line with human observations. The KinD method significantly increases the lightness level, but local texture features are poorly handled, such as blurred license plate numbers and large area artifacts in the sky. Other methods have obvious redness or overexposure.

Fig. \ref{fig12} shows the enhancement results of a low-light image with noise. It can be seen from the cat's ear that HE and AIEMC overemphasize the local details and lose the features. The HE method hardly amplifies the noise in the shadow area on the right side of the cat's ear, while the other methods amplify the noise signal, the worst are NPEA, AIEMC, MF and KinD. For the KinD method, there are serious artifact defects near the grass, and the noise in the cat's ear shadow area is overamplified. The LIME method produces appropriate results. The overall color saturation is high, but there is still strong noise. The noise intensity of our method and EFF is low, and the noise signal is not overamplified.

The six sets of legends in Fig.\ref{fig8} are taken from the HDR, LOL and synthetic datasets, where images in normal light are regarded as ``ground truth'' and can be used to reveal the appropriateness of color and texture detail. For a series of results from image X$_{4}$, the proposed method and EFF are very close to reality. For image X$_{5}$, HE produces significant color distortion and loses some of the features of the ceiling lamp. KinD also loses the local texture of the ceiling lamp to make it no longer visible. NPEA and LIME also {\color{blue}produced} a discordant result, which excessively enhanced the ceiling lamp area. SRIE and KinD produce additional shadows in the suspension area. The remaining methods deal with the characteristics of the ceiling lamp area better. In addition, it can be seen that our method fully restores the details, and the lightness is appropriate. For image X$_{6}$, the result obtained by our method is very close to the reference image. Most of the contrast methods retain the color saturation of the sky in the low light image but some local textures are lost. HE and KinD achieve poor results. 

For image X$_{7}$, our results present milky white square boxes, which are different from the others. For image X$_{8}$, our results are close to those of normal images but the NPEA method appears to be superior. For image X$_{9}$, the lightness difference between the normal image and the test image is very small, but our method still achieves suitable results.

For the nine comparison methods, the lightness level is not an adjustable parameter, and there is no lightness relationship similar to Eq. \eqref{eq:eq11} in this paper. Eq. \eqref{eq:eq11} enables our system to control lightness levels. After setting a specified lightness level or lightness difference, the gamma value parameters can be automatically adjusted for different images based on Eq. \ref{eq:eq11}. The comparison method lacks the function of flexibly adjusting and controlling the lightness level so that the lightness level of the result becomes uncontrollable.

A series of image comparison results show that the proposed method has better adaptability to various complex low-light environments and can reveal the dark area details of the low-light image and reproduce natural colors.

%

\subsection{Objective Quantitative Examination}
In this section, we show the index scores of different images in the qualitative evaluation experiment. In addition, to objectively evaluate the performance of different algorithms, the HDR, LOL and synthetic datasets are used for objective and quantitative examination. These three datasets have reference images under normal lighting, and performance measurements using these datasets are more convincing than {\color{blue}using datasets} without reference objects (such as SIDD, ExDark and NASA). 

\begin{figure}[h]
	\centerline{\includegraphics[width=\columnwidth]{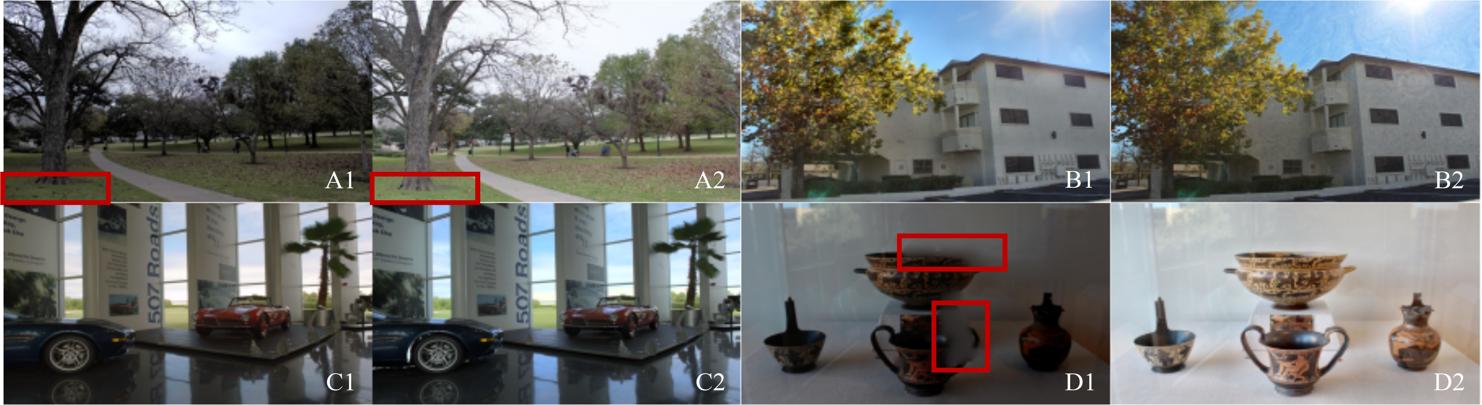}}
	\caption{{\color{blue}Four common defects in the HDR dataset. A: inconsistent size; B: artifact; C: only hue change; D: artificial cover mark.}}
	\label{fig16}
\end{figure}

{\color{blue}A total of} 250 pairs of images are selected from the HDR dataset to undertake the index evaluation experiment. The HDR dataset uses 605 scenes, totaling 1811 images. After carefully examining each image, we found that there are many pairs of images in the HDR dataset with inconsistent sizes, {\color{blue}artifacts and other defects}, as shown in Fig. \ref{fig16}. To improve the reliability of the experimental data and results, we manually browse through each image and eventually filter out 250 pairs of visually pleasing images. These 250 pairs of images have been packaged into a new modified dataset and published on the Internet. The LOL dataset contains 500 pairs of low/high light images, and the synthetic dataset contains 1000 pairs of low/high light images. Considering that the KinD algorithm has used the LOL dataset as the training set to obtain the network parameters, to fairly evaluate the performance of the algorithm, the index score of the LOL dataset does not include that of the KinD algorithm. The index evaluation experiment consists of 7 different measurement methods, which are derived from two different index systems of no-reference assessment and referential assessment. These are NIQE\cite{b49}, BRISQUE\cite{b50}, $\Delta E$, PSNR, MSSIM\cite{b40}, VIF\cite{b41} and LOE\cite{b46}. Among them, the smaller the index score is, the higher the image quality represented by NIQE, BRISQUE, $\Delta E$ and LOE. The other three indicators are the opposite.

NIQE and BRISQUE are two popular blind image quality assessment models, both of which are based on a statistical analysis of spatial domain characteristics to avail the evaluation score and depict the distortion intensity of the image. 

$\Delta E$ is a measure of color accuracy that represents the distance between two colors. This index needs to be calculated in the LAB color space. Specifically, we used the $\Delta E$ index to evaluate the color difference between the enhanced image and the reference image. If the index score is smaller, the color representing the enhanced image is closer to the reference image. The $\Delta E$ index is calculated from the following formula. 
\begin{equation}\label{eq:eq17}
\Delta E=\frac{1}{N_{I}} \sum_{i, j \in \Omega} \sqrt{\left(L_{i j}^{\prime}-L_{i j}\right)^{2}+\left(a_{i j}^{\prime}-a_{i j}\right)^{2}+\left(b_{i j}^{\prime}-b_{i j}\right)^{2}}
\end{equation}
where $N_{I}$ is the total number of pixels in an image of size $m \times n$. $(i, j)$ is the image coordinate number. $(L_{i j}^{\prime}, a_{i j}^{\prime}, b_{i j}^{\prime})$ and $(L_{i j}, a_{i j}, b_{i j})$ are the specific values of the enhanced image and the reference image in the LAB space at coordinates $(i, j)$, respectively.

PSNR, MSSIM, and VIF are three commonly used reference image quality assessment indexes. PSNR is a measure of the similarity between two images. MSSIM is usually used to describe the detailed reconstruction intensity of an image. VIF is the abbreviation of visual information fidelity, which depicts the degree of information fidelity. The LOE index is a measure of the lightness order between an enhanced image and a low-light image, which is used to evaluate the degree of lightness order preservation. 

\begin{figure}[h]
	\centerline{\includegraphics[width=0.95\columnwidth]{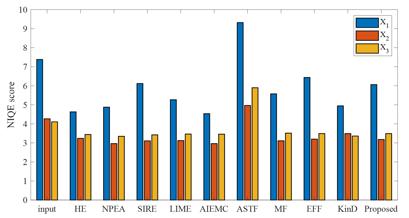}}
	\caption{The NIQE scores of different methods for no-reference images.}
	\label{fig13}
\end{figure}

\begin{figure}[h]
	\centerline{\includegraphics[width=0.95\columnwidth]{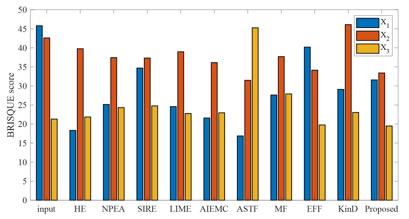}}
	\caption{The BRISQUE scores of different methods for no-reference images.}
	\label{fig14}
\end{figure}

Figs. \ref{fig13} and \ref{fig14} show the NIQE and BRISQUE scores of no-reference images X$_{1}$, X$_{2}$, and X$_{3}$, respectively. For image X$_{1}$, our method achieves a very pleasant visual effect but does not obtain the best score, which means that these two blind evaluation criteria do not always work well. For image X$_{2}$, our NIQE score is almost the same as any other method except ASTF, although our method yields suitable facial results. In addition, our method achieves almost the best BRISQUE scores on images X$_{2}$ and X$_{3}$.

\begin{table*}[!t]
	\caption{Average score of different enhancement algorithms on seven metrics for HDR datasets}
	\label{table2}
	\setlength{\tabcolsep}{3pt}
	\begin{tabular}{p{34pt}<{\centering}p{34pt}<{\centering}p{34pt}<{\centering}p{34pt}<{\centering}p{34pt}<{\centering}p{34pt}<{\centering}p{34pt}<{\centering}p{34pt}<{\centering}p{34pt}<{\centering}p{34pt}<{\centering}p{34pt}<{\centering}p{34pt}<{\centering}p{34pt}<{\centering}}
		\hline Metric & Low & Normal & HE & NPEA & SRIE & LIME & AIEMC & ASTF & MF & EFF & KinD & Proposed \\
		\hline NIQE$\downarrow$ & 3.3748 & 3.0299 & 3.2215 & 2.9901 & 3.1833 & 3.0768 & 3.1766 & 5.2234 & 3.0973 & 2.9841 & 3.7504 & \textbf{2.9696} \\
		\hline BRISQUE$\downarrow$ & 19.3845 & 17.0602 & 21.5078 & 18.1562 & 20.1574 & 18.9236 & 19.2683 & 39.6097 & 18.7311 & 17.8536 & 27.7495 & \textbf{16.7143} \\
		\hline $\Delta E$$\downarrow$ & 27.0268 & 0 & 21.1646 & 17.3746 & 18.1377 & 18.0564 & 18.1277 & 17.9176 & 15.7433 & 14.5737 & 17.8647 & \textbf{14.1556} \\
		\hline PSNR$\uparrow$ & 11.9782 & - & 15.0982 & 16.3342 & 15.7121 & 16.2825 & 16.7529 & 15.4505 & 17.4332 & 17.7794 & 16.1883 &\textbf{18.2541} \\
		\hline MSSIM$\uparrow$ & 0.5895 & - & 0.7592 & 0.8109 & 0.7918 & 0.8104 & 0.8234 & 0.7774 & 0.8366 & 0.8548 & 0.8042 & \textbf{0.8558} \\
		\hline VIF$\uparrow$ & 0.4254 & - & 0.4894 & 0.5047 & 0.5242 & \textbf{0.5643} & 0.4882 & 0.4139 & 0.5185 & 0.5354 & 0.4483 & 0.5099 \\
		\hline LOE$\downarrow$ & 0 & 785.9825 & 774.6608 & 404.4183 & 175.5405 & 1054.2414 & \textbf{72.8435} & 795.7680 & 380.0537 & 287.0465 & 701.0843 & 143.1179 \\
		\hline
	\end{tabular}
	\label{tab2}
\end{table*}

\begin{table*}[!t]
	\caption{Average score of different enhancement algorithms on seven metrics for LOL datasets}
	\label{table5}
	\setlength{\tabcolsep}{3pt}
	\begin{tabular}{p{37pt}<{\centering}p{37pt}<{\centering}p{37pt}<{\centering}p{37pt}<{\centering}p{37pt}<{\centering}p{37pt}<{\centering}p{37pt}<{\centering}p{37pt}<{\centering}p{37pt}<{\centering}p{37pt}<{\centering}p{37pt}<{\centering}p{37pt}<{\centering}}
		\hline Metric & Low & Normal & HE & NPEA & SRIE & LIME & AIEMC & ASTF & MF & EFF & Proposed \\
		\hline NIQE$\downarrow$ & 7.2452 & 4.7837 & 8.9739 & 8.6564 & \textbf{7.4947} & 8.4568 & 8.7162 & 13.8490 & 9.1068 & 7.7608 & 7.9986 \\
		\hline BRISQUE$\downarrow$ & 17.0079 & 18.8372 & 45.0078 & 41.0106 & 25.7858 & 36.7214 & 36.5421 & 64.7310 & 41.8458 & \textbf{24.5617} & 31.0244 \\
		\hline $\Delta E$$\downarrow$ & 44.9838 & 0 & 30.4496 & 23.3047 & 30.4476 & 21.7687 & 22.6604 & 31.2560 & 22.0068 & 24.8376 & \textbf{21.6160} \\
		\hline PSNR$\uparrow$ & 7.7433 & - & 14.1168 & \textbf{15.9962} & 11.4163 & 15.9631 & 15.5722 & 11.0340 & 15.8631 & 13.4927 & 15.8680 \\
		\hline MSSIM$\uparrow$ & 0.2054 & - & 0.6349 & 0.7290 & 0.5907 & 0.7321 & 0.7285 & 0.5683 & 0.7458 & 0.7141 & \textbf{0.7833} \\
		\hline VIF$\uparrow$ & 0.2130 & - & 0.3452 & 0.3691 & 0.3622 & \textbf{0.4011} & 0.3554 & 0.3008 & 0.3565 & 0.3656 & 0.3879 \\
		\hline LOE$\downarrow$ & 0 & 351.7104 & 779.4965 & 501.9875 & 90.1751 & 391.5140 & \textbf{9.5909} & 630.4621 & 201.4941 & 143.8986 & 188.7948 \\
		\hline
	\end{tabular}
	\label{tab5}
\end{table*}

\begin{table*}[!t]
	\caption{Average score of different enhancement algorithms on seven metrics for synthetic datasets}
	\label{table6}
	\setlength{\tabcolsep}{3pt}
	\begin{tabular}{p{34pt}<{\centering}p{34pt}<{\centering}p{34pt}<{\centering}p{34pt}<{\centering}p{34pt}<{\centering}p{34pt}<{\centering}p{34pt}<{\centering}p{34pt}<{\centering}p{34pt}<{\centering}p{34pt}<{\centering}p{34pt}<{\centering}p{34pt}<{\centering}p{34pt}<{\centering}}
		\hline Metric & Low & Normal & HE & NPEA & SRIE & LIME & AIEMC & ASTF & MF & EFF & KinD & Proposed \\
		\hline NIQE$\downarrow$ & 4.8169 & 4.1817 & 4.8809& 4.3935 & 4.3614  & 4.6950 & 4.7905 & 7.8054 & 4.4200 & \textbf{4.2962} & 4.6613 & 4.3970 \\
		\hline BRISQUE$\downarrow$ & 22.6080 & 17.2668 & 24.6007 & 19.1697 & 20.4624 & 23.0934 & 21.8838 & 47.4482 & 20.4482 & \textbf{18.6455} & 28.2160 & 18.7024 \\
		\hline $\Delta E$$\downarrow$ & 33.1859 & 0 & 19.1127 & 18.5786 & 23.7615 & 16.6366 & \textbf{14.5805} & 21.6365 & 18.0799 & 19.8329 & 18.2926 & 17.2697 \\
		\hline PSNR$\uparrow$ & 11.6074 & - & 16.0801 & 16.4299 & 14.7857 & 16.8888 & \textbf{18.5019} & 14.7833 & 17.2739 & 16.9172 & 16.8772 &18.1326 \\
		\hline MSSIM$\uparrow$ & 0.5220 & - & 0.7886 & 0.8214 & 0.7125 & 0.8075 & 0.8629 & 0.7685 & 0.8189 & 0.8167 & 0.8450 & \textbf{0.8676} \\
		\hline VIF$\uparrow$ & 0.4179 & - & 0.5627 & 0.5875 & 0.5441 & \textbf{0.6200} & 0.5965 & 0.4784 & 0.6159 & 0.5912 & 0.4936 & 0.5741 \\
		\hline LOE$\downarrow$ & 0 & 153.0237 & 305.5002 & 221.2730 & 120.5617 & 513.7948 & 82.9887 & 440.6012 & 155.4508 & 136.4731 & 317.7432 & \textbf{67.2860} \\
		\hline
	\end{tabular}
	\label{tab6}
\end{table*}

\begin{table*}[!t]
	\caption{Time cost of different enhancement algorithms under different size images (time in seconds)}
	\label{table4}
	\setlength{\tabcolsep}{3pt}
	\begin{tabular}{p{115pt}<{\centering}p{34pt}<{\centering}p{34pt}<{\centering}p{34pt}<{\centering}p{34pt}<{\centering}p{34pt}<{\centering}p{34pt}<{\centering}p{34pt}<{\centering}p{34pt}<{\centering}p{34pt}<{\centering}p{34pt}<{\centering}}
		\hline Image Size & HE & NPEA & SRIE & LIME & AIEMC & ASTF & MF & EFF & KinD & Proposed \\
		\hline 5312$\times$2988$\times$3 & 4.1094 & 467.5722 & 427.5963 & 7.3335 & 4.0926 & 4.6450 & 10.6938 & 7.4460 & - & \textbf{1.8111} \\
		\hline 2000$\times$1312$\times$3 & 0.6659 & 76.5653 & 133.6892 & 1.2267 & 0.7423 & 0.7446 & 1.6453 & 1.3929 & - & \textbf{0.2876} \\
		\hline 960$\times$540$\times$3 & 0.1372 & 15.2183 & 19.5285 & 0.2059 & 0.1752 & 0.1454 & 0.4393 & 0.2631 & 0.3406 & \textbf{0.0688} \\
		\hline 500$\times$375$\times$3 & 0.0477 & 5.4250 & 8.3621 & 0.0725 & 0.0778 & 0.0550 & 0.1866 & 0.1038 & 0.1469 &  \textbf{0.0330} \\
		\hline
	\end{tabular}
	\label{tab4}
\end{table*}

{\color{blue}Tables} \ref{tab2}, \ref{tab5} and \ref{tab6} show the scores of different methods under seven indicators and show the quality assessment of low-light images (briefly expressed as Low) and normal light images (briefly expressed as Normal), but these two types of results are only for display. Bold fonts are used to highlight the top algorithms in the tables. 

For the HDR dataset, the proposed method achieves the best performance for the five indicators and is only slightly inferior for the VIF and LOE indicators. 

For the LOL dataset, our method has the highest scores on the $\Delta E$ and MSSIM metrics and {\color{blue}ranks} third in the NIQE, BRISQUE and PSNR metrics. For this dataset, the VIF and LOE metrics are slightly lower.
	
For the synthetic dataset, our method obtains the highest scores on the MSSIM and LOE metrics, {\color{blue}ranks} second in the BRISQUE and PSNR metrics, and {\color{blue}ranks} third in the $\Delta E$ metric. For this dataset, the VIF and NIQE metrics are slightly lower.

Our method achieves the two best performances on the $\Delta E$ index, which indicates the superiority of the proposed algorithm in color restoration. This is also consistent with the results of the subjective evaluation experiments mentioned above. Our method has the best performance in three rounds of tests on the MSSIM index, which shows the superiority of the proposed algorithm in detail recovery. The LIME method has the best performance in three rounds of tests on the VIF index, but its LOE index is huge, which indicates that the lightness order has been greatly changed. The AIEMC method achieves the two best {\color{blue}performances} on the LOE index, which means that the lightness order of most enhancement results approaches the reference image.

Combining the results of the three datasets, we conclude that our method is very competitive.

\subsection{Time Cost}
Another important measure of algorithm performance is the time cost. Table \ref{tab4} records the time cost of different enhancement algorithms at different picture sizes. The time cost was obtained by averaging the time of 10 records. We used a GPU to speed up the calculation for the KinD method, while the other methods use a CPU. Constrained by the video memory of the GPU, we cannot complete the time statistic test of the KinD method when the image size is 2000$\times$1312$\times$3 or above. As shown in Table \ref{tab4}, the time cost of the proposed algorithm is the lowest, and the time consumption is reduced by at least half. The EFF method is close to the proposed method on {\color{blue}the seven indexes}. However, our method is at least 4$\times$ more efficient than the EFF method for the images with a size of 2000$\times$1312$\times$3 and more, which means that real-time processing of video with a conventional image size is entirely possible. We implemented the algorithm using MATLAB 2018b, so it can be further accelerated by using faster programming languages, such as C/C++.

\section{Conclusion}
\label{sec:conc}
In this paper, we have proposed a simple yet effective low-light image enhancement method to perceive unknown information {\color{blue}from dark areas}. This is the first application of the proposed cell vibration model in the field of image vision. Based on the analysis of the energy model, we have improved the standard gamma model, that is, we {\color{blue}retained} more image features with a large increase in lightness. After further investigation, the statistical {\color{blue}properties} revealed a relationship between natural image lightness and gamma intensity. To further optimize the local details of global lightness enhancement images, we proposed a local fusion strategy that successfully recovers the nature of the original low-light images. Comprehensive experiments have been undertaken and
The results show that our algorithm {\color{blue}recovered image details well and effectively avoided overenhancement} and color distortion. Compared with several {\color{blue}state-of-the-art} methods, our method achieves a step forward in image enhancement tasks. In future work, we will continue to optimize the performance and time consumption of the algorithm, hoping to {\color{blue}maintain} systematic efficiency even for large-scale images of 100 million pixels.


%

\appendices
\section{Equation derivation}
{\color{blue}This appendix mainly describes the mathematical derivation process of Eqs. \eqref{eq:eq6} and \eqref{eq:eq8} in subsection B.}

The {\color{blue}stimulation} energy equations we derived are as follows. To apply these two equations to the image field, we need to rewrite the equations.
\begin{equation}\label{eq:eq26}
	\varepsilon_{s}=c_{1} c_{2} \sqrt{k M}-\frac{1}{2} c_{2}^{2} M
\end{equation}
\begin{equation}\label{eq:eq27}
	E=\frac{1}{2 \pi} c_{1} c_{2} k-\frac{1}{4 \pi} c_{2}^{2}\sqrt{k M}
\end{equation}

Assume
\begin{equation}\label{eq:eq28}
	\left\{\begin{array}{c}
		\lambda=c_{1} c_{2} \sqrt{k} \\
		1-\lambda=-\frac{1}{2} c_{2}^{2}
	\end{array}\right.
\end{equation}

Due to the smoothness of Eq. \eqref{eq:eq26}, the above assumptions ensure that $\varepsilon_{s}(M=0)=0, \varepsilon_{s}(M=1)=1$. We can derive
\begin{equation}\label{eq:eq29}
	\left\{\begin{array}{c}
		k=\frac{\lambda^{2}}{c_{1}^{2} c_{2}^{2}} \\
		c_{2}=\sqrt{2(\lambda-1)}
	\end{array}\right.
\end{equation}

Substitute the above parametric relationships into Eq. \eqref{eq:eq27},
\begin{equation}\label{eq:eq30}
	\begin{aligned}
		E&=\frac{1}{2 \pi} c_{1} c_{2} k-\frac{1}{4 \pi} c_{2}^{2} \sqrt{k M} \\
		 &=\frac{1}{2 \pi} c_{1} c_{2} \frac{\lambda^{2}}{c_{1}^{2} c_{2}^{2}}-\frac{1}{4 \pi} c_{2}^{2} \frac{\lambda}{c_{1} c_{2}} \sqrt{M} \\
		 &=\frac{\lambda^{2}}{2 \pi c_{1} c_{2}}-\frac{\lambda c_{2}}{4 \pi c_{1}} \sqrt{M} \\
		 &=\frac{\lambda^{2}}{2 \pi c_{1} \sqrt{2(\lambda-1)}}-\frac{\lambda \sqrt{2(\lambda-1) M}}{4 \pi c_{1}} \\
		 &=\frac{1}{2 \sqrt{2} \pi c_{1}}\left[\frac{\lambda^{2}}{\sqrt{\lambda-1}}-\lambda \sqrt{(\lambda-1) M}\right]
	\end{aligned}
\end{equation}

To make Eq. \eqref{eq:eq30} more {\color{blue}concisely}, we assume
\begin{equation}\label{eq:eq31}
	c_{1}=\frac{1}{2 \sqrt{2} \pi}
\end{equation}

{\color{blue}Therefore}, we can obtain
\begin{equation}\label{eq:eq32}
	E=\frac{\lambda^{2}}{\sqrt{\lambda-1}}-\lambda \sqrt{(\lambda-1) M}
\end{equation}

It is assumed that the pixel value of an image represents the stimulus intensity, that is, $M = I$, so
\begin{equation}\label{eq:eq33}
	\varepsilon_{s}=\lambda \sqrt{I}+(1-\lambda) I
\end{equation}
\begin{equation}\label{eq:eq34}
	E=\frac{\lambda^{2}}{\sqrt{\lambda-1}}-\lambda \sqrt{(\lambda-1) I}
\end{equation}


%
%

\ifCLASSOPTIONcaptionsoff
  \newpage
\fi



%

%
\begin{IEEEbiography}[{\includegraphics[width=1in,height=1.25in, clip,keepaspectratio]{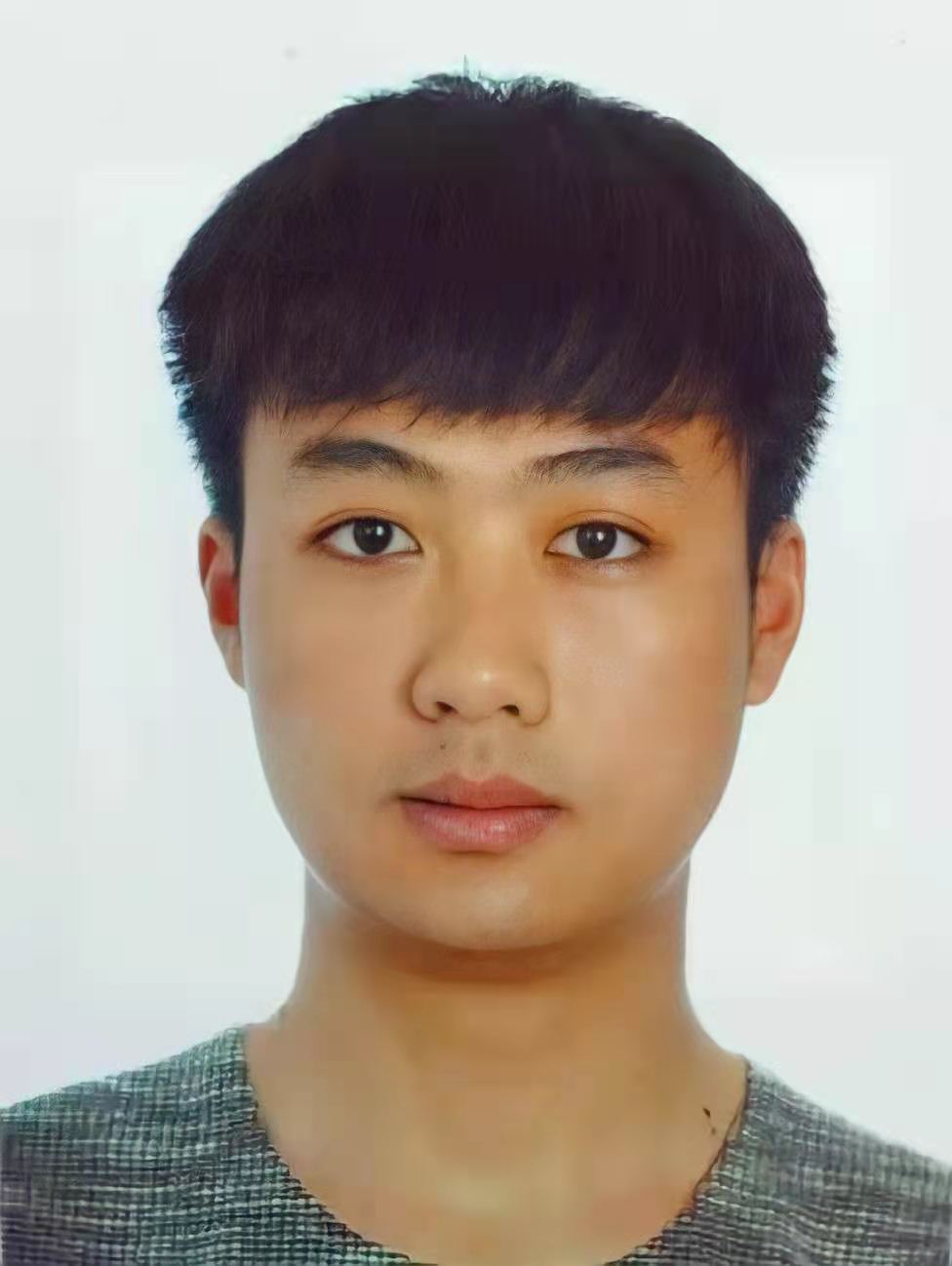}}]{Xiaozhou Lei} received a B.S. degree in mechanical design manufacture and automation major from Wuhan Institute of Technology, Wuhan, China, in 2015. and an M.S. degree in mechatronic engineering from Wuhan Institute of Technology, Wuhan, China, in 2018. He is currently pursuing a Ph.D. degree in control science and engineering at Shanghai University, Shanghai, China. His current research interests include machine vision, artificial intelligence, image enhancement, and object detection.
\end{IEEEbiography}

\begin{IEEEbiography}[{\includegraphics[width=1in,height=1.25in, clip,keepaspectratio]{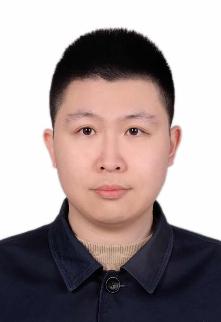}}]{Zixiang Fei} received his bachelor's degree at Liverpool John Moores University and his master's degree at the University of York. He received his PhD degree in 2020 and worked as a postdoc in 2021 at the University of Strathclyde. Now, he is a lecturer at Shanghai University. His major research interests include computer vision, machine learning, object recognition and deep learning.
\end{IEEEbiography}

\begin{IEEEbiography}[{\includegraphics[width=1in,height=1.25in, clip,keepaspectratio]{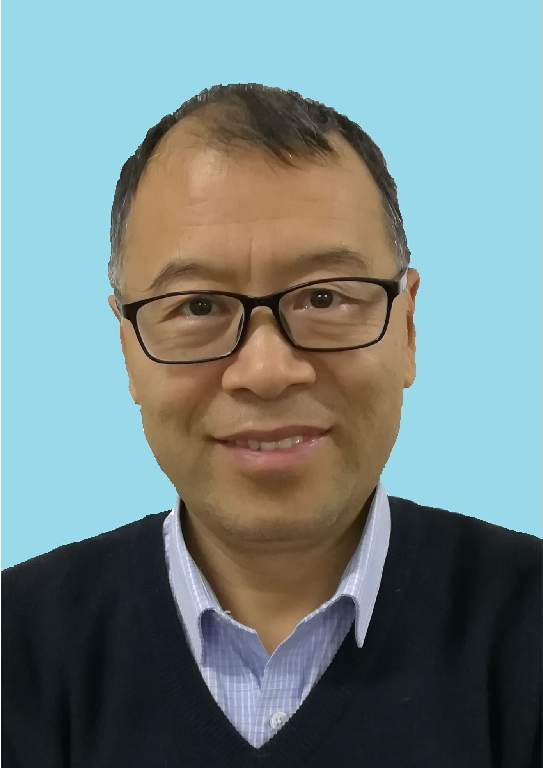}}]{Wenju Zhou} received a B.Sc. and M.Sc. degrees from Shandong Normal University, China in 1990 and 2005, and his Ph.D. degree from Shanghai University, China in 2014. He is now a distinguished researcher and Doctoral Supervisor at Shanghai University. His research interests include robotics control, machine vision, and the industry applications of automation equipment.
\end{IEEEbiography}

\begin{IEEEbiography}[{\includegraphics[width=1in,height=1.25in, clip,keepaspectratio]{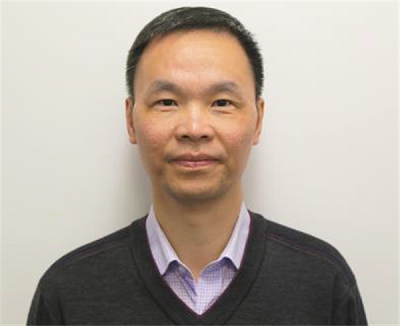}}]{Huiyu Zhou} received his B.E. degree in radio technology from Huazhong University of Science and Technology, China, an M.Sc. degree in biomedical engineering from the University of Dundee, U.K., and the D.Phil. degree in computer vision from Heriot-Watt University, Edinburgh, U.K. He is currently a Professor with the School of Computing and Mathematical Sciences, University of Leicester, United Kingdom. He has published widely in the field.
\end{IEEEbiography}

\begin{IEEEbiography}[{\includegraphics[width=1in,height=1.25in, clip,keepaspectratio]{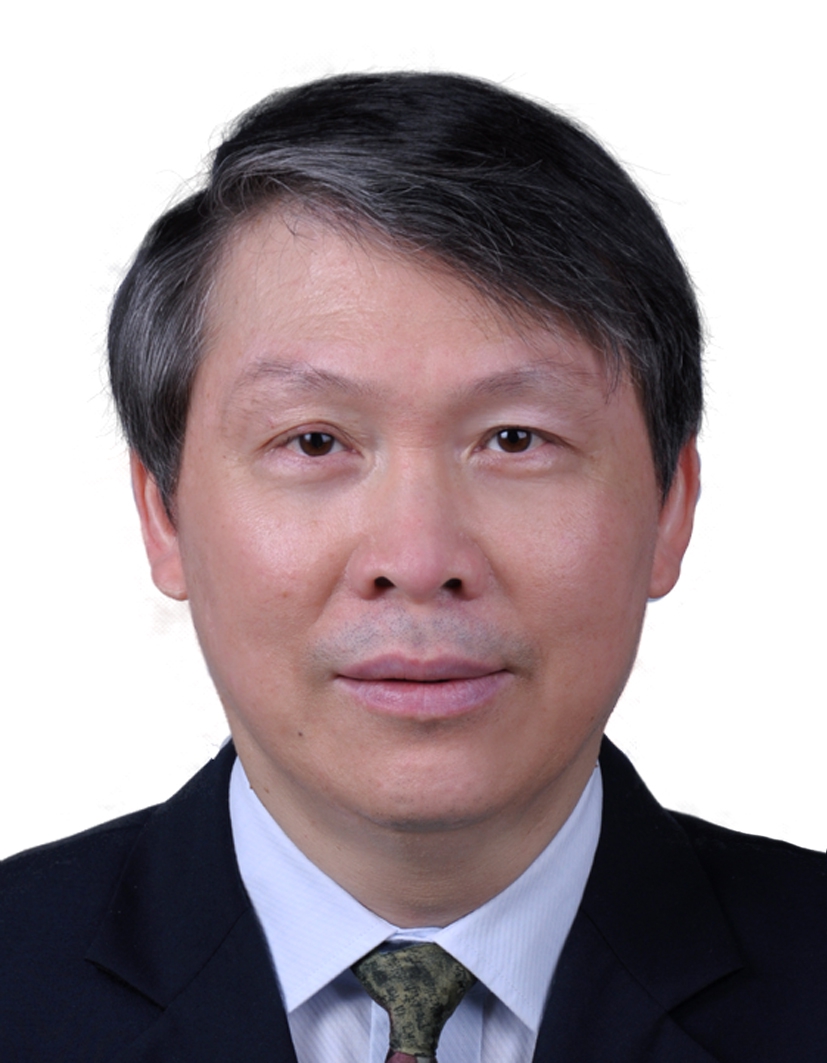}}]{Minrui Fei} received his B.S. and M.S. degrees in Industrial Automation from the Shanghai University of Technology in 1984 and 1992, respectively, and his PhD degree in Control Theory and Control Engineering from Shanghai University in 1997. Since 1998, he has been a full professor at Shanghai University. He is Chairman of Embedded Instrument and System Subsociety, and Standing Director of China Instrument \& Control Society; Chairman of Life System Modeling and Simulation Subsociety, Vice-chairman of Intelligent Control and Intelligent Management Subsociety, and Director of Chinese Artificial Intelligence Association, Fellow of China Simulation Federation. His research interests are in the areas of networked control systems, machine vision, artificial intelligence, intelligent control, complex system modeling, hybrid network systems, and field control systems.
\end{IEEEbiography}




\end{document}